\documentclass[twocolumn,ip]{jpsj3}
\usepackage{txfonts}
\title{
Transport Properties and Diamagnetism of Dirac Electrons in Bismuth}

\author{Yuki Fuseya$^1$\thanks{E-mail: fuseya@pc.uec.ac.jp}, 
Masao Ogata$^2$, 
and 
Hidetoshi Fukuyama$^3$}
\inst{$^1$Department of Engineering Science, University of Electro-Communications, Chofu, Tokyo 182-8585 \\
$^2$Department of Physics, University of Tokyo, 7-3-1 Hongo, Bunkyo-ku, Tokyo 113-0033\\
$^3$Department of Applied Physics and Research Institute for Science and Technology, Tokyo University of Science, Kagurazaka, Shinjuku-ku, Tokyo 162-860} 

\abst{
	Bismuth crystal is known for its remarkable properties resulting from particular electronic states, e. g., the Shubnikov-de Haas effect and the de Haas-van Alphen effect.
	Above all, the large diamagnetism of bismuth had been a long-standing puzzle soon after the establishment of quantum mechanics, which had been resolved eventually in 1970 based on the effective Hamiltonian derived by Wolff as due to the interband effects of a magnetic field in the presence of a large spin-orbit interaction. 
	This Hamiltonian is essentially the same as the Dirac Hamiltonian, but with spatial anisotropy and an effective velocity much smaller than the light velocity.
	This paper reviews recent progress in the theoretical understanding of transport and optical properties, such as the weak-field Hall effect together with the spin Hall effect, and ac conductivity, of a system described by the Wolff Hamiltonian and its isotropic version with a special interest of exploring possible relationship with orbital magnetism.
	It is shown that there exist a fundamental relationship between spin Hall conductivity and orbital susceptibility in the insulating state on one hand, and the possibility of fully spin-polarized electric current in magneto-optics.
	Experimental tests of these interesting features have been proposed.
	}

\usepackage{amsmath,bm,mathrsfs,color}
\renewcommand{\Im}{{\rm i}}

\newcommand{\D}{\Delta}

\newcommand{\bk}{\bm{k}}
\newcommand{\bp}{\bm{p}}
\newcommand{\bq}{\bm{q}}
\newcommand{\br}{\bm{r}}

\newcommand{\bpi}{\bm{\pi}}
\newcommand{\bsgm}{\bm{\sigma}}
\newcommand{\kp}{k\cdot p}
\newcommand{\bW}{\bm{W}}

\newcommand{\wc}{\omega_{\rm c}^*}
\newcommand{\mc}{m_{\rm c}^*}

\newcommand{\scr}[1]{\mathscr{#1}}
\newcommand{\ve}{\varepsilon}
\newcommand{\tve}{\tilde{\varepsilon}}
\newcommand{\sgn}{{\rm sgn}}
\newcommand{\RH}{R_{\rm H}}

\begin{document}
\maketitle

\section{Introduction}\label{Sec_Intro}
	
	Bismuth has played an important role in solid state physics.\cite{Wilson_text,Mott-Jones_text,Peierls_text}
	Many key phenomena are discovered firstly in bismuth (Table \ref{List2}).
	These discoveries have elucidated by the remarkable properties of bismuth, such as low carrier densities, small effective masses, high mobilities, long mean free path and large $g$-factor (Table \ref{List1}).

	The anomalously large diamagnetism\cite{Faraday1845,Curie1895} and highly efficient thermoelectricity\cite{Seebeck1822,Ettingshausen1886} have been realized in the 19 th century.
	Its electrical transport phenomena were examined repeatedly; the highest Hall coefficient\cite{Onnes1912} and the highest magnetoresistance\cite{Kapitza1928} were reported in bismuth early in the 20 th century.
	The discoveries that deserve special attention are the Schubnikov-de Haas\cite{Schubnikov1930} and the de Haas-van Alphen\cite{Haas1930} effects in 1930.
	At that time both quantum oscillations were quite mysterious.
	It was Peierls who first gave a quantitative theory of this oscillations based on the Landau's quantum theory of diamagnetism\cite{Peierls1933b}.
	%
\begin{table}[tb]
\caption{List of phenomena discovered first in bismuth.}
\label{List2}
\begin{tabular}{ll}
\hline
Year & Discovery  \\
\hline
1778 & diamagnetism (Brugmans; named by Faraday in 1845)\\
1821 & Seebeck effect \\
1886 & Nernst effect (Ettingshausen \& Nernst)\\
1928 & Kapitza's law of magnetoresistance\\
1930 & Shubnikov-de Haas effect\\
1930 & de Haas-van Alph\'en effect\\
1955 & cyclotron resonance in metals (Galt)\\
1963 & oscillatory magnetostriction (Green \& Chandrasekhar)\\
\hline
\end{tabular}

\caption{List of specific properties of bismuth mainly for electrons at $L$-points. ($n_{\rm e}$ and $n_{\rm h}$ are the electron and hole carrier densities, respectively. $m$ being the bare electron mass.)}
\label{List1}
\begin{tabular}{ll}
\hline
low carrier densities & $n_{\rm e} = n_{\rm h}\sim 10^{17} {\rm cm}^{-3}$\\
small effective masses  & $\mc \sim 10^{-3} m$\\
high mobilities & $\mu \sim 10^8 {\rm cm}^2 {\rm V}^{-1}{\rm s}^{-1}$\\
long mean free path &  $\ell \sim 0.3$mm\\
large diamagnetism & $ \chi \sim -10^{-5} {\rm emu}$\\
large $g$-factor & $ g\sim 1000$\\
\hline
\end{tabular}

\end{table}
	%
	With rapid growth of semiconductor physics, bismuth was studied more intensively by newly developed experimental and theoretical techniques.
	Various experimental methods, such as the cyclotron resonance\cite{Galt1955,Galt1959,Lax1960b}, the electron spin resonance\cite{Smith1960,Everett1962}, and the magneto-infrared reflection\cite{Lax1958,Boyle1960a}, which are all developed for the semiconductor physics, were applied also to bismuth. 
	For the analysis of these experimental results, the effective model Hamiltonian was introduced\cite{Cohen1960,Lax1960,Wolff1964}.
	It was shown by Wolff\cite{Wolff1964} that the effective Hamiltonian for bismuth, the Wolff Hamiltonian, is essentially equivalent to the Dirac Hamiltonian, but with spatial anisotropy of effective velocity.
	Hence, electrons in bismuth began to be called ``Dirac electrons".
	Up to the 1970s, the electronic structure of bismuth has been clarified in detail; bismuth is one of the best understood materials similar to silicon and germanium.\cite{Golin1968,Ferreira1967,Ferreira1968,Dresselhaus1971,Edelman1976,Issi1979}
	Then the subjects of interest gradually shifted from the electronic structure toward the anomalous properties of bismuth.
	The topics are rich in variety: the diamagnetism\cite{Adams1953,Fukuyama1970,Otake1980}, the magneto-optics in the quantum limit\cite{Maltz1970,Vecchi1974,Vecchi1976,Edelman1975,Hiruma1983}, the thermoelectricity\cite{Smith1962,Gallo1963,Lenoir1996,Lenoir2001,Dresselhaus2003,Goldsmid_text}, the excitonic insulator,\cite{Mott1961,Halperin1968a,Halperin1968b,Mase1971,Fukuyama1971b,Fukuyama1971c,Nakajima1976,Yoshioka1976a,Yoshioka1976b,Kuramoto1979} etc.

\begin{figure}[t]
	\begin{center}
		\includegraphics[width=5cm,bb=0 0 188 251]{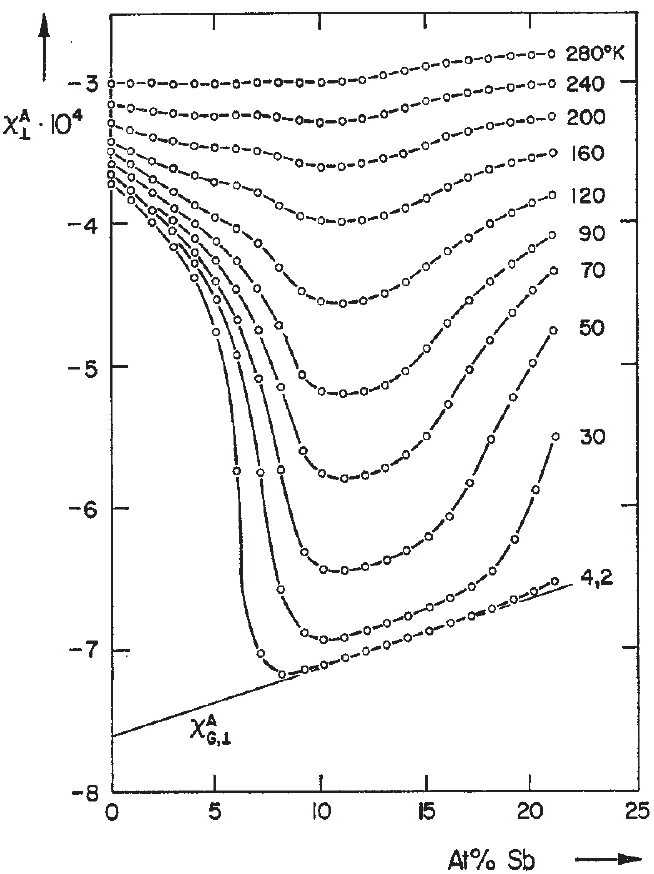}
	\end{center}
\caption{Magnetic susceptibility per gram of Bi$_{1-x}$Sb$_x$ at different temperatures for the field perpendicular to the trigonal axis.
	The chemical potential locates in the band gap for $7\% \lesssim x \lesssim 20 \%$. Taken from Ref. \citen{Wehrli1968}.
}
\label{Wehrli}
\end{figure}

	Among these, the diamagnetism is one of the most important phenomena in bismuth.
	The extremely large diamagnetism of bismuth cannot be explained based on the Landau-Peierls formula,\cite{Peierls1933} which is the standard theory of diamagnetism.
	The Landau-Peierls formula is derived by an approximation for a particular Bloch band and by neglecting the interband matrix elements of a magnetic field.
	It predicts that the orbital susceptibility is proportional to the density of state. 
	On the other hand, even early experiments of diamagnetism on bismuth and its alloys in 1930's indicate that the diamagnetism takes its maximum value when the chemical potential is located in the band gap\cite{Goetz1934,Shoenberg1936a,Jones1934,Mott-Jones_text}, i.e., the insulating state as is shown more clearly in the experiment of Wehrli in 1968\cite{Wehrli1968} as shown in Fig. \ref{Wehrli}.
	%
	 %
	 This was a long-standing problem that many physicist tried to understand.
	 The mystery was finally solved when the following points are taken into account\cite{Fukuyama1970}: (i) the interband effect of a magnetic field, (ii) the large spin-orbit interaction, and (iii) the specific band structure of bismuth based on the Wolff Hamiltonian. 

	The transport phenomena are more complex. 
	The ordinary transport phenomena, such as the longitudinal electric conductivity, are basically dissipative, whereas the diamagnetic current is dissipationless.
	By contrast, the Hall effect, which is commonly believed to be also dissipative, is quite special, since the dissipationless diamagnetic current should also play some roles in the presence of a magnetic field.
	It was proposed that the Hall conductivity and the orbital magnetism should be related to each other in some way through the interband effect of a magnetic field\cite{Kubo1970}, although the details of the relation had been unknown.

	Recently, the interband effects on the transport phenomena has been investigated in detail for Dirac electrons in solids.\cite{Fukuyama2007,Kobayashi2008,Fuseya2009,Fuseya2012a,Fuseya2012b,Fukuyama2012,Fuseya2014}
	It has been revealed that the interband effect of a magnetic field gives rise to an unconventional contribution to the Hall conductivity, which is remarkable at the band-edge and almost independent from the impurity scatterings; these properties are common features with that of the diamagnetism.
	Furthermore, it has been shown that the finite spin Hall conductivity, which is realized by the interband contributions, exists even in the insulating states, and is related to the orbital susceptibility by a simple formula only with the physical constants.\cite{Fuseya2012b,Fuseya2014}
	The interband effect can also induce the fully spin-polarized electric current by using the circularly polarized light.\cite{Fuseya2012a}
	These recent progresses have been made through the study based on the Dirac Hamiltonian in solids.
	In this paper, we shall give a review of the recent progress on the transport phenomena and its relation to the diamagnetism of Dirac electrons in solids, especially in bismuth.

	We shall first describe the general properties of bismuth.
	The crystal structure and the electron energy spectrum are given in \S\ref{Crystals}.
	After the brief explanation of the $\kp$ theory, the effective Hamiltonian of bismuth, the Wolff Hamiltonian, is introduced in \S\ref{Sec_Dirac}, where some general properties of Dirac electrons in solids are given.
	Also, some general properties under a magnetic field are given in \S\ref{Sec_magnetic}.
	The second part of this review deals the specific properties of Dirac electrons in solids.
	The diamagnetism, which is one of the most distinctive property of Dirac electrons, are reviewed in \S\ref{Sec_Diamagnetism}, paying special attention to the interband effect of a magnetic field.
	The interband effect on the dc transport phenomena is discussed for the Hall effect in \S\ref{Sec_Hall}.
	Another interesting phenomena, the spin Hall effect, is argued in \S\ref{Sec_SHE} for the Dirac Hamiltonian in solids.
	The spin Hall effect is further studied for the Wolff Hamiltonian and the quantitative evaluation of the spin Hall conductivity for bismuth are given in \S\ref{Aniso}.
	In \S\ref{Sec_SPEC}, a 100\% spin-polarized electric current is proposed.
	Section \ref{Sec_Summary} is devoted for the summary.
	%

\section{Crystal Structure and Electron Energy Spectrum}\label{Crystals}

	Bismuth, antimony and arsenic are the group V semimetals.
	Their electron energy spectrum share many properties in common, and is responsible for interesting similarities between them.
	The narrow band gap located in the vicinity of the Fermi level dominates the physical properties, such as the small effective mass, high mobilities, and non-parabolic dispersion.
	This energy spectrum can be controlled by alloying with another element or applying pressure.

\subsection{Crystal structure}
\begin{figure}
\begin{center}
\includegraphics[width=8cm,bb=0 0 311 156]{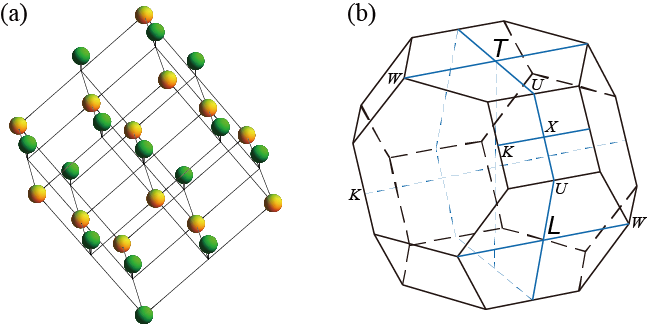}
\end{center}
\caption{(Color online) (a) Crystal structure of group V semimetals. This structure is obtained from the simple cubic lattice by stretching the lattice along the body diagonal direction, and by shifting the atoms in one sublattice (indicated by green) along the body diagonal direction.
(b) Brillouine zone of group V semimetals. It is similar to the Brillouine zone of fcc lattice, but slightly shrinked due to the lattice distortion; $T$-point and $L$-point are not equivalent.
}
\label{Crystal}
\end{figure}
\begin{figure}[tb]
\begin{center}
\includegraphics[width=8cm,bb=0 0 458 217]{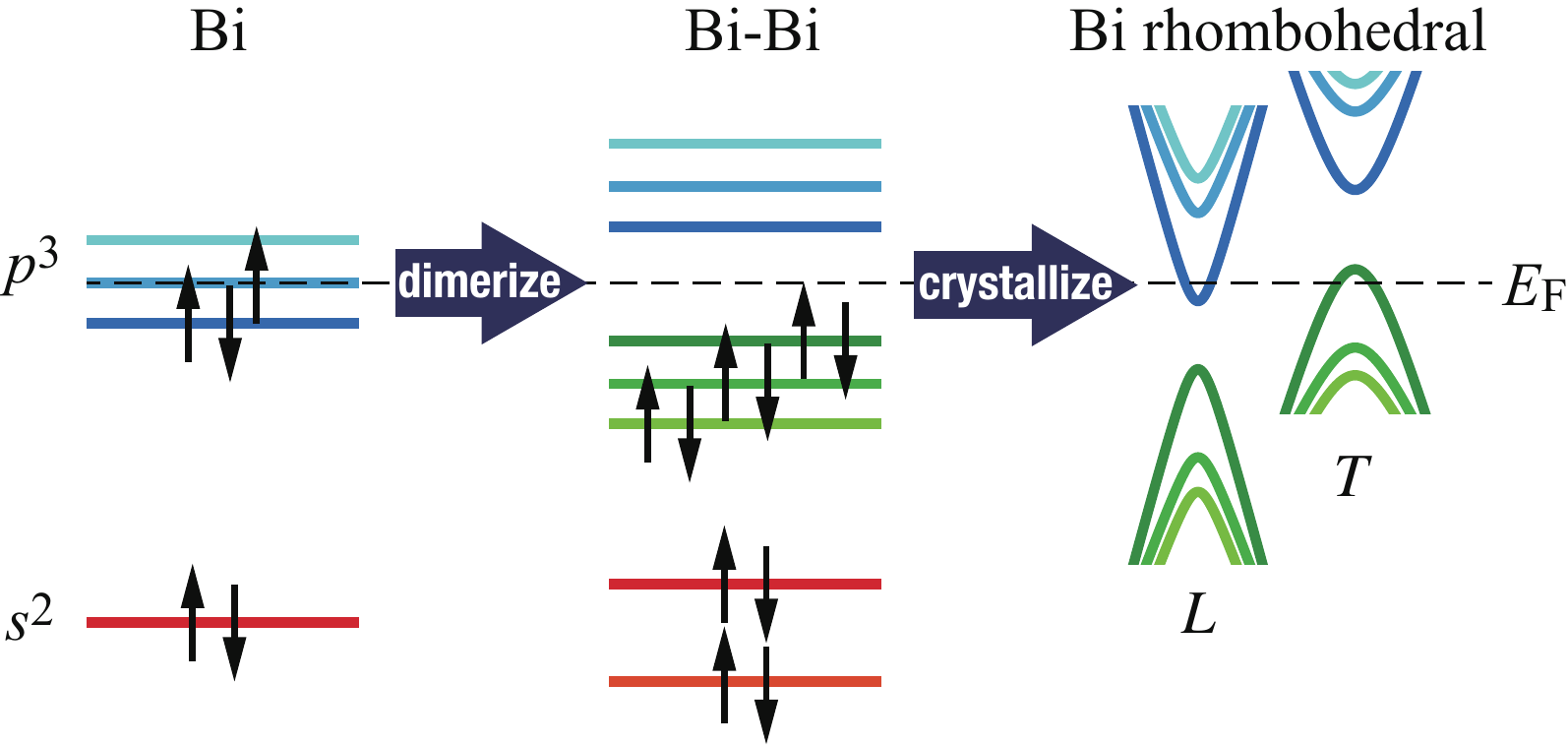}
\end{center}
\caption{(Color online) Schematic picture of the energy spectrum of bismuth (group V elements).
For the single atom, the $p$-band is half-filled.
By the dimerization, an energy gap opens, which lowers the total energy: the system becomes insulator.
With a small lattice distortion, the conduction and valence bands are hybridized in order to gain the kinetic energy of carriers and lower the total energy further.
}
\label{schematic}
\end{figure}

	%
	The energy spectrum mentioned above is the reflection of the crystal structure of group V elements. 
	The group V elements crystallize in the rhombohedral structure (the so-called arsenic or A7 structure) as shown in Fig. \ref{Crystal} (a).
	Basic features of the rhombohedral structure can be understood as follows (see Fig. \ref{schematic}).
	The group V elements have odd-number ($s^2p^3$) electrons, so that they should be metallic.
	However, they lower the energy by forming the dimarization, i.e., the Peierls distortion\cite{Peierls_text}.
	There are two atoms in the unit cell by this distortion and there are even-number (10 valence) electrons, so that the system can be insulating.
	As a matter of fact, since this lattice distortion is very weak and then the gap is very small, the energy spectrum can be either a narrow-gap semiconductor or a semimetal. 
	Pure As, Sb and Bi crystals favor the semimetallic energy spectrum, while Bi with 7-20 \% Sb tends to be a narrow-gap semiconductor.
	%
	%
	Note that for the IV-VI compounds, where there are ten valence electrons in the unit cell of the rock-salt structure, the dimerization is stronger than group V elements due to the stronger ionization, so that the system favors the insulating state.
	%
	
	The followings are the details of the crystal structure of group V elements.
	It originates from the two interpenetrating fcc lattices like the rock-salt structure.
	If the origin of one sublattice is taken at $(0, 0, 0)$, that of the other sublattice is taken at $(2u, 2u, 2u)$.
	For the undistorted rock-salt structure, $u=1/4$, and the rhombohedral angle $\alpha$, which is the angle between the unit vectors, is $\alpha = 60^\circ$.
	The rhombohedral structure is obtained by the following two kind of distortions:
	\begin{itemize}
		\item shift the location of atoms in one sublattice relative to the other along the body diagonal $(1 1 1)$ direction,
		\item stretch the both sublattice along the $(1 1 1)$ direction.
\end{itemize}
	The former modifies $u$ from $1/4$, and the latter changes $\alpha$ from $60^\circ$.
	(The initial length of the unit vectors are kept.)
	%
	%
	%
	%
	%
	The rhombohedral structure so obtained loses many symmetries from the simple cubic. 
	The parameters for the group V semimetals and some IV-VI compounds are summarized in Table. \ref{t3}.
\begin{table}
\caption{Crystal structure parameters of the group V semimetals and some IV-VI compounds.\cite{Cohen1964}
Note that PbS, PbSe and PbTe are all semiconductors.}
\label{t3}
\begin{center}
\begin{tabular}{lll}
\hline
 & $\alpha$ & $u$ \\
\hline
As	& $54^\circ 10'$ & 0.226\\
Sb	& $57^\circ 6.5'$ & 0.233\\
Bi	& $57^\circ 14.2'$ & 0.237\\
PbS	& $60^\circ$ & 0.25\\
PbSe& $60^\circ$ & 0.25\\
PbTe	& $60^\circ $ & 0.25\\
\hline
\end{tabular}
\end{center}
\end{table}
	%
	
	The Brilouin zone for the group V semimetals is shown in Fig. \ref{Crystal} (b).
	It is given by squeezing the Brillouin zone of fcc lattice, the truncated octahedron, along the trigonal direction.
	The high symmetry points are labeled similarly to that for the fcc lattice.
	The exception is the $L$-points. 
	The original $L$-points of the fcc lattice change their symmetry: the two points of higher symmetry are labeled by $T$, and the remaining six equivalent points with lower symmetry are labeled by $L$.

	There are one trigonal axis along $\Gamma T$ direction, three binary axes along $TW$, three bisectrix axes along $TU$.
	Usually, binary, bisectrix and trigonal axes are denoted by $x$, $y$ and $z$ or $1$, $2$ and $3$, respectively.
	%

\subsection{Electron energy spectrum of bismuth}\label{Sec_spectrum}
\begin{figure}[tb]
\begin{center}
\includegraphics[width=5cm,bb=0 0 148 241]{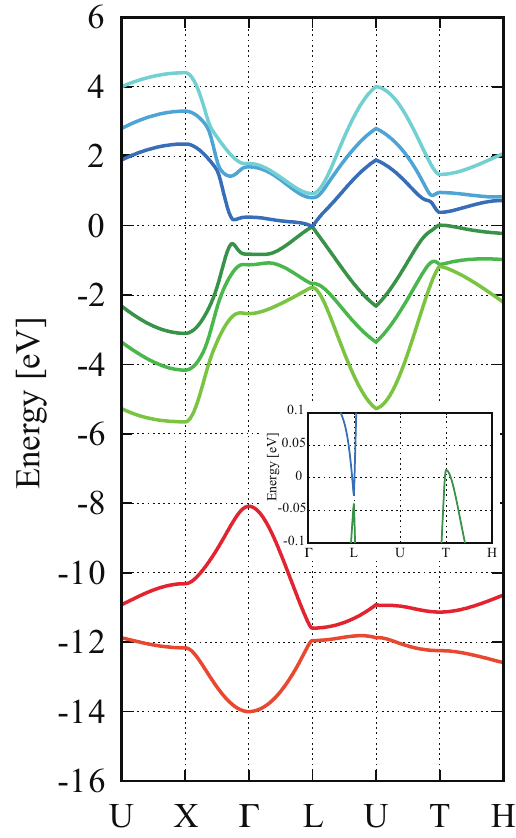}
\end{center}
\caption{(Color online) Band structure of bismuth by the tight-binding calculation.\cite{Liu1995}
}
\label{band}
\end{figure}
	
	%
	Various first principle calculations for bismuth have been carried out so far (e.g., the pseudo potential approach\cite{Golin1968}, the APW method\cite{Ferreira1967,Ferreira1968} and the pseudo potential DFT\cite{Gonze1990}), but the accuracy is not enough.
	(The required accuracy is the energy scale of the band gap $\sim 15$meV.)
	The energy dispersion obtained by Liu and Allen with the tight-binding calculation\cite{Liu1995} is shown in Fig. \ref{band}.
	This is one of the most reliable band calculation for bismuth.
	Liu and Allen considered up to the third-neighbor bonding with the on-site spin-orbit interaction, and adjust the parameters in order to fit with the experimental results.
	They determined the parameters in order to reproduce (i) the overlap between the highest valence and lowest conduction bands, (ii) the Fermi energy, (iii) the effective masses, (iv) the shapes of the Fermi surfaces, and (v) the band gaps near the Fermi level.
	Their results give good agreements with experimental results.

	Recently, the angle-resolved Landau spectrum measurements with high accuracy have carried out,\cite{LLi2008,HYang2010,ZZhu2011,ZZhu2012} and the electronic energy spectrum have been determined in more great detail.\cite{Sharlai2009,Alicea2009}
	The ``extended" Wolff Hamiltonian, which takes into account the contributions from higher energy bands in addition to the two band near the Fermi level (considered by the Wolff Hamiltonian, cf. \S\ref{Sec_Dirac}), was newly devised based on the $\kp$ theory in order to fit the recent experimental progress\cite{ZZhu2011,ZZhu2012}; this gives the best fitting for the recent experimental results including the spin splittings.

	The effective mass of bismuth so obtained is quite small.
	For example, the cyclotron mass of the electron at $L$-point of bismuth is $\mc = 0.00189$ for $\bm{H}\, ||$ bisectrix axis.\cite{ZZhu2011}
	The small cyclotron mass leads to the large g-factor, ($g^* = 1060$ for $\mc = 0.00189$), since the g-factor of the Dirac electron is given by the inverse of the cyclotron mass.
	The large spin magnetic-moment accompanied by the large g-factor can generates large spin-responses, such as the spin Hall effect or the spin-polarized electric current, which are the main topics of this review.

\section{Effective Hamiltonian}\label{Sec_Dirac}
\subsection{$\kp$ theory in the Luttinger-Kohn representation}\label{kp}

	Here we introduce the method of Luttinger and Kohn\cite{Luttinger1955}, which clearly represents the essence of the $\kp$ theory.
	Let $\scr{H}$ be the Hamiltonian of the electron in a periodic potential with spin-orbit interaction as
	\begin{align}
		\scr{H}=\frac{p^2}{2m}+V(\br)+\frac{\hbar}{4m^2 c^2} \bsgm \cdot \bm{\nabla}V(\br)\times \bp,
		\label{init}
\end{align}
	where $m$ is the bare electron mass, $c$ the velocity of light, $\bsgm$ the Pauli spin matrix vector, $V(\br)$ being the periodic potential arising from the crystal, and $\bp$ is the momentum operator $-\Im \hbar \bm{\nabla}$.
	Since $V(\br)$ is periodic and $\bp$ is invariant under translations, the eigenfunction of the above Hamiltonian will have the form of Bloch functions as
	\begin{align}
		\psi_{n\bk}(\br)=e^{\Im \bk \cdot \br}u_{n\bk}(\br).
		\label{Bloch}
\end{align}
	The $u_{n\bk}$ are spin-dependent periodic functions.
	Of course, the $u_{n\bk}$ form a complete set of functions, in which any wave function can be expanded.
	Now we consider the set of functions
	\begin{align}
		\chi_{n\bk}(\br)=e^{\Im \bk \cdot \br}u_{n\bk_{0}}(\br),
\end{align}
	where we assume that there is an extremum of the energy band at $\bk_{0}$.
	The functions $\chi_{n\bk}$ also form a complete orthogonal set, so that any wave function can be expanded \emph{rigorously} in terms of $\chi_{n\bk}$.
	The orthogonality of $\chi_{n \bk}$ also holds:
	\begin{align}
		\langle \chi_{n \bk} | \chi_{n' \bk'} \rangle = \delta(\bk'-\bk)\delta_{nn'}.
	\end{align}
	The $u_{n\bk}$ can be expanded in terms of $u_{n \bk_{0}}$ as
	\begin{align}
		u_{n\bk}(\br)=\sum_{n} b_{nn'}(\bk)u_{n'\bk_{0}}(\br).
		\label{utrans}
\end{align}
	The eigenfunction of Schr\"{o}dinger equation $\psi$ is expanded in terms of $\chi_{n\bk}$ as
	\begin{align}
		\psi (\br) = \sum_{n} \int\!\! d\bk \, c_{n}(\bk) \chi_{n\bk}, 
		\label{eq6}
\end{align}
	which gives the equation
	\begin{align}
		\sum_{n'}\int\!\! d\bk' \, \langle n\bk |\scr{H}|n'\bk'\rangle c_{n'}(\bk')=\epsilon c_{n}(\bk).
		\label{eqA}
\end{align}
	Here the matrix elements with respect to $\chi_{n\bk}$ are evaluated as
	\begin{align}
		 \langle n\bk |\scr{H}|n'\bk'\rangle 
		 &= \int \!\! d\br \, e^{\Im (\bk'-\bk)\cdot \br} u_{n\bk_{0}}^{*}
		 \left( \epsilon_{n' \bk_{0}} + \frac{ \hbar \bk'\cdot \bp}{m} + \frac{\hbar^2 k'^{2}}{2m} \right) u_{n' \bk_{0}},
\end{align}
	where $\epsilon_{n \bk_{0}}$ is the energy at the extremum of the $n$-th band.
	Considering the orthonormality of $\chi_{n\bk}$, we have
	\begin{align}
		\langle n\bk |\scr{H}|n'\bk'\rangle
		&=\delta(\bk-\bk')
		\left[
			\left( \epsilon_{n\bk_{0}}+\frac{\hbar^2 k^{2}}{2m} \right) \delta_{nn'}
			+ \frac{\hbar \bk \cdot \bp_{nn'}}{m}
		\right],
\end{align}
	where
	\begin{align}
		\bp_{nn'}=\frac{(2\pi)^{3}}{\Omega}\int_{\rm cell} \!\! d\br \,u_{n\bk_{0}}^{*}
		\left\{ \bp + \frac{\hbar}{4mc^{2}}\bsgm \times \bm{\nabla}V(\br) \right\}
		u_{n\bk_{0}},
\end{align}
	with $\Omega$ the volume of the unit cell.
	Then we obtain the eigenvalue equation of $\kp$ theory from eq. (\ref{eqA}) in the form
	\begin{align}
		\sum_{n'}\left[
		\left( \epsilon_{n \bk_{0}} + \frac{\hbar^2 k^2}{2m}\right) \delta_{nn'} + \frac{\hbar \bk \cdot \bm{p}_{nn'} }{m}
		\right]
		c_{n'}(\bk)
		=\epsilon c_{n}(\bk).
		\label{eqFin}
\end{align}
	The matrix elements of the momentum operator should satisfy $\bm{p}_{nn}/m=0$, since we assume that the band has an extremum at $\bk_{0}$. 
	%

\subsection{Wolff Hamiltonian}
	
	The $\kp$ theory is rigorous as long as we consider the complete set of $\chi_{n\bk}$.
	In semimetals and semiconductors, the physical properties are dominated by the narrow region in the $\bk$-space around the extremum.
	In such a situation, just a few $\chi_{n\bk}$ can give a quantitatively good approximation.
	For example, in the case of bismuth, it has been shown that only two bands are enough for understanding the experimental results.\cite{Dresselhaus1971}

	Cohen and Blount applied the $\kp$ theory to the two band model with the spin-orbit interaction in order to express the low energy properties of electrons at $L$-points in bismuth.\cite{Cohen1960}
	In this case, the eigenfunction $\psi$ is expanded by four $\chi_{n\bk}$'s (conduction and valence bands with up and down spins), so that eq. (\ref{eqFin}) becomes
	\begin{align}
		\begin{pmatrix}
			\D & 0 & \hbar \bk \cdot \bm{t} & \hbar \bk \cdot \bm{u}\\
			0 & \D & -\hbar \bk \cdot \bm{u}^* & \hbar \bk \cdot \bm{t}^*\\
			\hbar \bk \cdot \bm{t}^* & -\hbar \bk \cdot \bm{u} & -\D & 0\\
			\hbar \bk \cdot \bm{u}^* & \hbar \bk \cdot \bm{t} & 0 & -\D
		\end{pmatrix}
		\begin{pmatrix}
			c_{1}\\
			c_{2}\\
			c_{3}\\
			c_{4}
		\end{pmatrix}
		=E
		\begin{pmatrix}
			c_{1}\\
			c_{2}\\
			c_{3}\\
			c_{4}
		\end{pmatrix}.
		\label{CB}
\end{align}
	Here, $\epsilon_{1 \bk_0}=\epsilon_{2 \bk_0}=\D$ and $\epsilon_{3 \bk_0}=\epsilon_{4 \bk_0}=-\D $, i.e., the band gap  is $2\D$, and $\bk$ is measured from $\bk_{0}$.
	Since the quadratic term $\hbar^2 k^2/2m$ in eq. (\ref{eqFin}) is relatively very small in bismuth, so that it is discarded in $E$ hereafter.
	The matrix elements of the velocity operator is given by $\bm{v}_{nn'}=\bm{p}_{nn'}/m$.
	Both $\bm{v}_{11}$ and $\bm{v}_{22}$ are zero, and also $\bm{v}_{12, 21}=\bm{v}_{34,43}=0$ due to the symmetry of the wave function for the crystal with a center of inversion.\cite{Yafet1963,note4}
	From the time-reversal and parity symmetry, we also have\cite{Cohen1960,note5}
	\begin{align}
		\bm{v}_{13}&=\bm{v}_{42}\equiv \bm{t},\\
		\bm{v}_{14}&=-\bm{v}_{32}\equiv \bm{u}.
\end{align}
	The Hamiltonian given by eq. (\ref{CB}) includes the case without the spin-orbit interaction, where only ${\rm Re}(\bm{t})$ is finite and ${\rm Im}(\bm{t})=0$, $\bm{u}=0$.
	(Orbital magnetism in the two-band model without the spin-orbit interaction is studied in Ref. \citen{Fukuyama1969}.)
	In the following, we discuss the case with a large spin-orbit interaction as $|\bm{u}|\simeq |\bm{t}|$.

	The four vectors, ${\rm Re}(\bm{t})$, ${\rm Im}(\bm{t})$, ${\rm Re}(\bm{u})$ and ${\rm Im}(\bm{u})$, are required to specify the two band model.
	However, one of these vectors can be eliminated by a suitable choice of the basis functions.
	Wolff chose these functions in such a way that ${\rm Re}(\bm{t})=0$.
	Then he found that the effective Hamiltonian of Cohen-Blount can be written in a very simple form as
	\begin{align}
	\scr{H}=
	\D \beta + \Im \hbar \bk \cdot \left[ \sum_{\mu=1}^3 \bW(\mu)\beta \alpha_{\mu} \right],
	\label{Wolff}
\end{align}
	where the vectors $\bW(\mu)$ are given by
	\begin{align}
		\bW(1) &= {\rm Im}(\bm{u}),\\
		\bW(2) &= {\rm Re}(\bm{u}),\\
		\bW(3) &= {\rm Im}(\bm{t}),
\end{align}
	and the $4\times 4$ matrices $\alpha_{\mu}$ and $\beta $ are 
	\begin{align}
		\alpha_{\mu}=
		\begin{pmatrix}
			0 & \sigma_{\mu}\\
			\sigma_{\mu} & 0
		\end{pmatrix},
		\quad
		\beta = 
		\begin{pmatrix}
			I & 0 \\
			0 & -I
		\end{pmatrix}.
\end{align}
	Equation (\ref{Wolff}) may be called Wolff Hamiltonian.
	The eigenenergy of this Hamiltonian is obtained as
	\begin{align}
		E = \pm \sqrt{\D^2 + \sum_{\mu=1}^3 \left[ \hbar \bk \cdot \bW(\mu) \right]^2}.
\end{align}

	It should be emphasized here that the Wolff Hamiltonian (\ref{Wolff}) is quite general.
	No particular properties of bismuth has been considered except the assumption $|\bm{u}|\simeq|\bm{t}|$.
	In other words, all the system that satisfies the following conditions can be expressed generally in terms of the Wolff Hamiltonian:
	\begin{enumerate}
		\item Time-reversal and parity symmetries are kept.
		\item Pair of conduction and valence bands are  isolated from the other bands.
		\item Band gap is much smaller than the other energy scales.
		\item Spin-orbit interaction is strong.
\end{enumerate}
	The characteristics of each material are reflected by $\D$ and $\bW (\mu)$, which are related to the inverse mass-tensor $\alpha_{ij}$ by the form
	\begin{align}
		\alpha_{ij}=\frac{1}{\D}\sum_{\mu} W_i (\mu) W_j (\mu). \label{inverse}
\end{align}
	Thus the third condition corresponds to the situation, where the effective mass is much smaller than the bare electron mass.

\begin{figure}[tb]
\begin{center}
\includegraphics[width=8cm,bb=0 0 322 147]{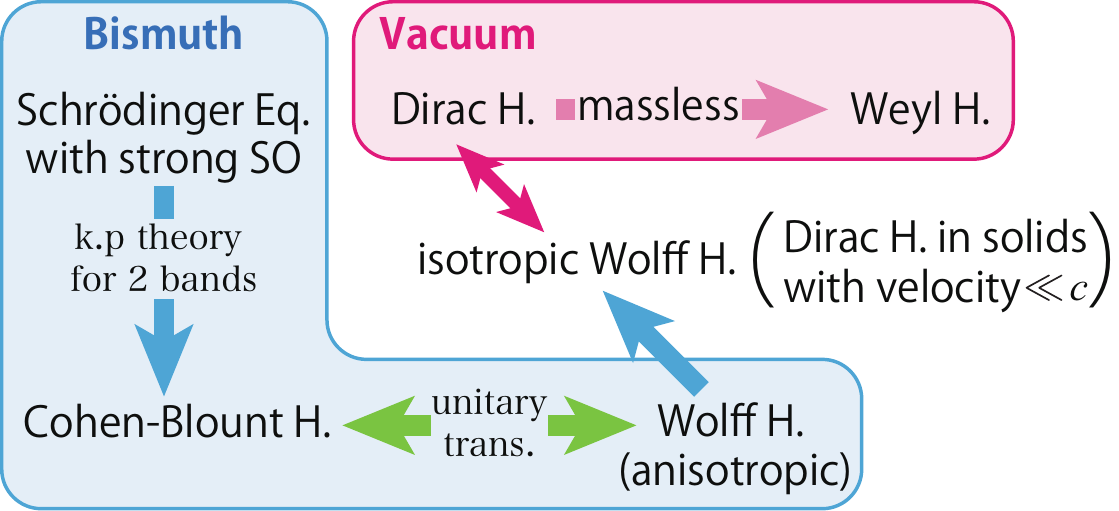}
\end{center}
\caption{(Color online) Relationship between the Wolff and Dirac Hamiltonian. (Here ``H." is an abbreviation for Hamiltonian.)
The application of the $\kp$ theory to the Schr\"{o}dinger equation with the strong spin-orbit (SO) interaction for two bands leads to the Cohen-Blount Hamiltonian and then, by a suitable choice of the basis, to the Wolff Hamiltonian, where the velocity is anisotropic in general.
When the velocity of the Wolff Hamiltonian is assumed to be isotropic, we obtain the Hamiltonian that is equivalent to the Dirac Hamiltonian but with the effective velocity $\gamma$ much less than the light velocity $c$.
This may be called the Dirac Hamiltonian in solids.
If the mass of Dirac Hamiltonian, which is written in terms of $4\times4$ matrix, is zero, we obtain the Weyl Hamiltoinian, which is written in terms of $2\times2$ matrix.
}
\label{DW}
\end{figure}

	When we assume the velocity vectors in the form
	\begin{align}
		\bW(1) &= (\gamma, 0, 0), \label{eq16}\\
		\bW(2) &= (0, \gamma, 0), \label{eq17}\\
		\bW(3) &= (0, 0, \gamma), \label{eq18}
\end{align}
	we have the isotropic Wolff Hamiltonian\cite{Fuseya2009,Fuseya2012a,Fuseya2012b}
	\begin{align}
		\scr{H}_{\rm iso}=
		\begin{pmatrix}
			\D & \Im \hbar  \gamma \bk \cdot \bsgm \\
			-\Im  \hbar \gamma \bk \cdot \bsgm & -\D
		\end{pmatrix},
		\label{isoWolff}
\end{align}
	which is essentially equivalent to the Dirac Hamiltonian\cite{Dirac1928},
	\begin{align}
		\scr{H}_{\rm D} =
		\begin{pmatrix}
			mc^2 & c \bp \cdot \bsgm \\
			c \bp \cdot \bsgm & -mc^2
		\end{pmatrix},
		\label{original}
\end{align}
	though the velocity in solids $\gamma$ is much smaller than the velocity of light $c$.
	The isotropic Wolff Hamiltonian keeps the essence of the original Wolff Hamiltonian, although the approximation eqs. (\ref{eq16})-(\ref{eq18}) may seem to be quite radical.
	Wolff was the first to point out the correspondence of the effective Hamiltonian in solids to the Dirac Hamiltonian.\cite{note1}
	In this review, we define ``Wolff Hamiltonian" by eq. (\ref{Wolff}) including the anisotropy, and ``isotropic Wolff Hamiltonian (or shortly ``Dirac Hamiltonian in solids"), we mean eq. (\ref{isoWolff}).
	The relation between Wolff and Dirac Hamiltonian is summarized in Fig. \ref{DW}.
	%
	%

\subsection{Dirac Hamiltonians of bismuth and graphene}
	
	Here we compare the above Dirac Hamiltonian of solids (\ref{isoWolff}), to the effective Hamiltonian of graphene given by:\cite{Slonczewski1958,Ando2005}
	\begin{align}
		\scr{H}_{\rm gr}=
		\begin{pmatrix}
			0 & \gamma (k_{x}-\Im k_{y})\\
			\gamma (k_{x}+\Im k_{y}) & 0
		\end{pmatrix}.
		\label{graphene}
\end{align}
	This Hamiltonian is two-dimensional Weyl Hamiltonian of massless particles.
	The effective Hamiltonian of $\alpha$-ET$_{2}$I$_{3}$ also belongs to this type.\cite{Kobayashi2007}
	The above Hamiltonian often expressed by using the Pauli matrix as $\scr{H}= \gamma \bk \cdot \bsgm$, which seems to be similar to eq. (\ref{isoWolff}).
	However, they are essentially different in the following points.
	First, the Dirac Hamiltonian of bismuth is written in terms of $4\times 4$ matrix, while that of graphene is in $2\times2$.
	Second, $\bsgm$ for bismuth expresses the degrees of freedom of real spins, while, in case of graphene, it expresses the degrees of freedom of the sublattice of the honeycomb lattice, namely, it has nothing to do with the real spins.
	The real-spin physics emerge in the Dirac Hamiltonian of bismuth, while the pseudo-spin physics appear in that of graphene.\cite{Koshino2010,Koshino2011}
	%

\section{Dirac electrons under a magnetic field}\label{Sec_magnetic}
\subsection{$\kp$ theory under a magnetic field}

	The result obtained by the $\kp$ theory, eq. (\ref{eqFin}), is equivalent to the result obtained by the Bloch representation, since they are related with each other by the unitary transformation eq. (\ref{utrans}).
	However, a significant difference appears in the calculation of physical quantities under a magnetic field. 
	Since the \emph{uniform} magnetic field is incompatible with the \emph{periodic} function $u_{n \bk}$, its treatment is extremely complex if based on the Bloch band.
	That is why the problem of the orbital susceptibility in a periodic potential was quite difficult to solve.
	On the other hand, in the $\kp$ theory, there is no $\bk$-dependence in the periodic function $u_{n \bk_{0}}$, which drastically reduces the complexity of the formulation.
	The applicability to the crystals under magnetic fields is one of the many successes of the $\kp$ theory.

	We consider the Hamiltonian (\ref{init}) under a magnetic field $\bm{H}$:
	\begin{align}
		\scr{H}=\frac{\left\{\bp+(e/c)\bm{A}\right\}^2}{2m}+V(\br)+\frac{\hbar}{4m^2 c^2} \bsgm \cdot \bm{\nabla}V(\br)
		\times \left\{\bp+(e/c)\bm{A}\right\},
		\label{init_mag}
\end{align}
	where $\bm{A}$ is a vector potential satisfying $\bm{H}=\bm{\nabla}\times \bm{A}$.
	We expand the eigenfunction in terms of $u_{n\bk_0}(\br)$ as (cf. eq. (\ref{eq6}))
	\begin{align}
		\psi(\br)= \sum_n \int\!\! d\bk c_n (\bk) e^{\Im \bk \cdot \br}u_{n\bk_0}(\br)=\sum_n F_n (\br) u_{n\bk_0}(\br).
		\label{env}
\end{align}
	%
	%
	Since the coefficients $F_n (\br)$ vary slowly in the scale of the lattice constant and they modulate the quickly oscillating lattice-periodic part $u_{n\bk_0}(\br)$, they are called envelope functions.
	We obtain the eigenvalue equation by substituting eq. (\ref{env}) into eq. (\ref{eqFin}) with a replacement $\hbar \bk \to \bpi \equiv -\Im \bm{\nabla} + (e/c)\bm{A}$ as\cite{Luttinger1955,Winkler_text,Voon_text}
	\begin{align}
		\sum_{n'}\left[
		\left( \epsilon_{n \bk_{0}} + \frac{\pi^2}{2m}\right) \delta_{nn'} + \frac{\bpi \cdot \bm{p}_{nn'} }{m}
		\right]
		F_{n'}(\br)
		=\epsilon F_{n}(\br).
\end{align}
	%
	%
	%
	It is straightforward to solve this equation, since all the information of $\epsilon_{\bk_{0}}$ and $\bm{p}_{nn'}$ are the same as that without magnetic field; 
	only the difference appears in the commutation relation $\bpi \times \bpi =-\Im (\hbar e/c) \bm{H}$.
	Furthermore, in the Luttinger-Kohn representation, it is clear how to keep the gauge invariance\cite{Fukuyama1969}, while it is very difficult (actually it is impossible in most cases) in the Bloch representation.
	This strong point of the $\kp$ theory for the gauge invariance is crucial for the trustworthy theory of transport and orbital magnetism.
	%

\subsection{Wolff Hamiltonian under a magnetic field}
	
	The Wolff Hamiltonian introduced in \S\ref{Sec_Dirac} is the result of the $\kp$ theory, so that the effect of the magnetic field is taken into account only by the replacement $\hbar \bk$ with $\bpi$ as
	\begin{align}
		\scr{H}
		\psi
		&= 
		\begin{pmatrix}
			\D & \Im \bpi\cdot \bm{\varLambda} \\
			-\Im \bpi \cdot \bm{\varLambda} & -\D
		\end{pmatrix}
		\psi
		=
		E\psi,
		\label{isoWolffmag}
\end{align}
	where 
	\begin{align}
		\bm{\varLambda}&=\sum_{\mu=1}^3 \bW (\mu) \sigma_\mu.
\end{align}
	In order to obtain the eigenvalues of this equation, we first consider the squared equation
	\begin{align}
		\scr{H}^2 
		\psi
		&=\begin{pmatrix}
		\D^2 + (\bpi\cdot\bm{\varLambda})^2 & 0\\
		0 & \D^2 +  (\bpi\cdot\bm{\varLambda})^2
		\end{pmatrix}
		\psi
		\nonumber\\
		&=
		\begin{pmatrix}
			\D^2 + 2\D \scr{H}^* & 0\\
			0 & \D^2 + 2\D \scr{H}^*
		\end{pmatrix}
		= E^2
		\psi,
		\\
		\scr{H}^* &= \frac{\bpi \cdot \hat{\alpha} \cdot \bpi}{2}
		+\bm{\mu}_{\rm s}^* \cdot \bm{H}.
		\label{Hstar}
\end{align}
	Here $\bm{\mu}_{\rm s}^*$ is the spin magnetic-moment near the extremum of band given by
	\begin{align}
		\bm{\mu}_{\rm s}^* 
		&=\frac{\hbar e\Omega}{2c\D} \sum_\mu \bm{Q}(\mu)\sigma_\mu,
		\\
		\bm{Q}(\lambda) &= \frac{\bW(\mu)\times \bW(\nu)}{\Omega}, \quad \mbox{($\lambda, \mu, \nu$ cyclic)}
		\\
		\Omega &=\bW(1)\times\bW(2) \cdot \bW(3).
\end{align}
	The second term of eq. (\ref{Hstar}), the Zeeman term, is not the original Zeeman term for the bare electrons, but is the effective Zeeman term resulting from the orbital motion of electrons, through the spin-orbit interaction.
	%
	%
	%
	%
	
	Since the first term of eq. (\ref{Hstar}), the orbital term, is the conventional form for the free electrons with an anisotropic effective mass, its eigenvalues are simply obtained as\cite{Ashcroft_text,Solyom_text}
	\begin{align}
		\hbar \wc \left( n + \frac{1}{2} \right) + \frac{\hbar^2 k_h^2}{2m_h},
\end{align}
	where $k_h$ is the wave vector along the magnetic field, the cyclotron frequency is given by
	\begin{align}
		\wc = \frac{eH}{\mc c},
\end{align}
	and the cyclotron effective mass $\mc$ is 
	\begin{align}
		\mc =\sqrt{\frac{\det \hat{m}}{m_h}}
\end{align}
	in which the longitudinal effective mass $m_h$ is
	\begin{align}
		m_h = \bm{h}\cdot \hat{m} \cdot \bm{h},
\end{align}
	with a unit vector along the magnetic field $\bm{h}$ and the mass tensor $\hat{m}=\hat{\alpha}^{-1}$.
	The eigenvalues of the second term of eq. (\ref{Hstar}), the Zeeman term, are obtained by considering its square
	\begin{align}
		\left( \bm{\mu}_{\rm s}^* \cdot \bm{H} \right)^2
		&=\left( \frac{\hbar e \Omega}{2c \D}\right)^2 \bm{H}\cdot \hat{A} \cdot \bm{H},
		\\
		A_{ij} &= \sum_\mu Q_i (\mu ) Q_j (\mu) 
		=\frac{\D^2}{\Omega^2} \frac{\hat{m}_{ij}}{\det \hat{m}}.
\end{align}
	Thus the eigenvalues of the Zeeman term are also written in terms of $\wc$ as
	\begin{align}
		\pm \frac{\hbar eH}{2c}\sqrt{\frac{m_h}{\det \hat{m}}}= \pm \frac{1}{2} \hbar \wc.
\end{align}
	The effective g-factor is determined from the effective Zeeman splitting $E_{\rm Z}^*$ through $g^* \mu_{\rm B}H = E_{\rm Z}^*$\cite{Yafet1963}, so that we obtain the effective g-factor
	\begin{align}
		g^* = \frac{2m}{\mc},
		\label{gfactor}
\end{align}
	where $\mu_{\rm B}=\hbar e/2mc$ is the Bohr magneton.
	The eigenvalues of $\scr{H}^*$ are then given as
	\begin{align}
		E^* = \hbar \wc \left(n+\frac{1}{2}\pm\frac{1}{2} \right) + \frac{\hbar^2 k_h^2}{2m_h}.
		\label{Estar}
\end{align}
	Finally, the eigenvalues of the Wolff Hamiltonian under a magnetic field are given by
	\begin{align}
		E=\pm \sqrt{\D^2 +2\D 
		\left\{
		\hbar \wc \left(n+\frac{1}{2}\pm\frac{1}{2} \right) + \frac{\hbar^2 k_h^2}{2m_h}
		\right\}}.
		\label{ene}
\end{align}
	%
	\begin{figure}
	\begin{center}
		\includegraphics[width=4.5cm,bb=0 0 321 283]{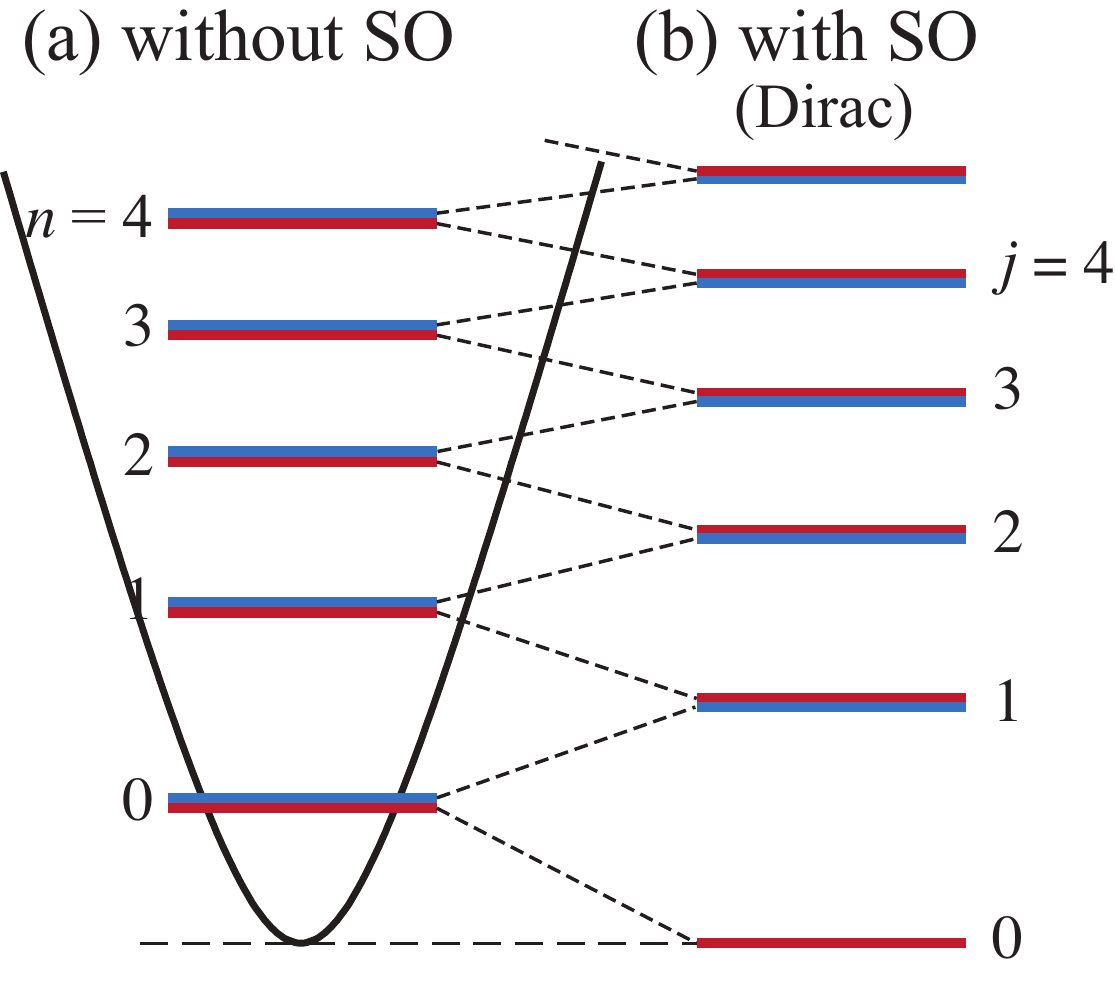}
	\end{center}
\caption{(Color online) Illustration of the energy levels under a magnetic field (a) without spin-orbit (SO) interaction, and (b) with SO interaction, i.e., Dirac electrons.
The horizontal lines indicate the position of energy levels with opposite spins.
We illustrate the $\kp$ theory for a two-band model without SO for (a)\cite{Lax1960}, where the energy level splitting is not constant.}
\label{models}
\end{figure}

	For free electrons, the Landau level splitting, $\hbar \omega_{\rm c}$, is the same as the Zeeman splitting $E_{\rm Z}$, and the quantized energy levels are labeled in terms of the orbital and the spin quantum numbers, which are independent quantum numbers.
	For electrons of the Wolff Hamiltonian, the Landau level splitting $\hbar \wc$ is also the same as $E_{\rm Z}^*$.
	However, the quantized energy levels are no more labeled in terms of the orbital and the spin quantum numbers, but are labeled in terms of the quantum number of total angular momentum:
	\begin{align}
		j=n+\frac{1}{2}\pm\frac{1}{2}=0, 1, 2, \ldots,
\end{align}
	which is the only good quantum number.
	This is due to the strong spin-orbit interaction, which completely mixes the orbital- and spin-angular momenta.

	The lowest energy level ($j=0$) is exceptional, since its spin state is uniquely determined; only the down spin of $n=0$ can occupy the lowest energy level.
	The $j=0$ level is fixed at the band-edge even in a strong magnetic field (Fig. \ref{models}). 
	This unique $j=0$ level enables the spin-polarized electric current discussed in \S\ref{Sec_SPEC}.
	%

\subsection{Anomalous diamagnetism}
	\begin{figure}
	\begin{center}
		\includegraphics[width=8cm,bb=0 0 644 834]{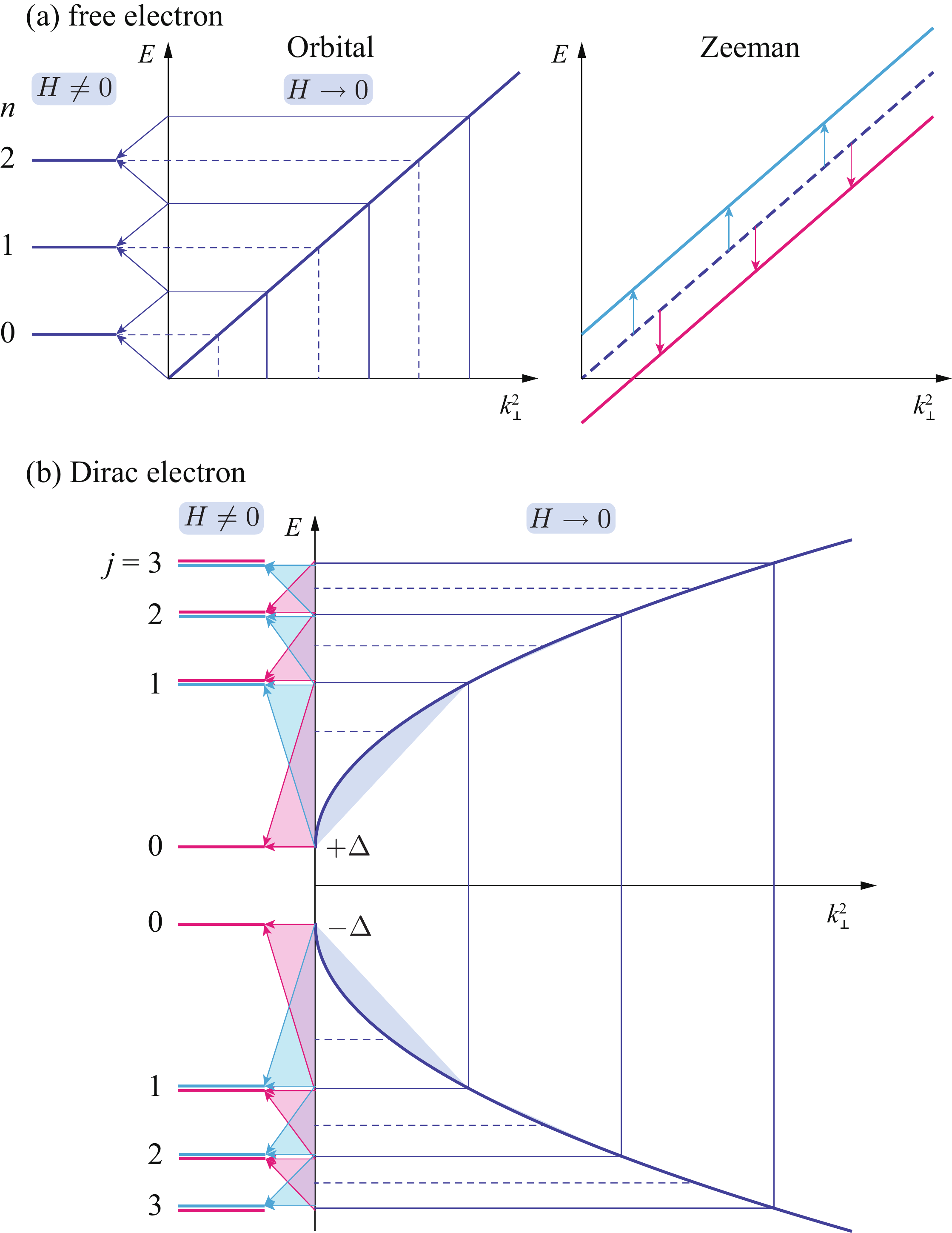}
	\end{center}
\caption{(Color online) Landau levels under the finite field, $H\neq 0$, and the energy for zero field limit, $H\to0$, as a function of $k_\perp^2$, where $k_\perp$ is the wave vector perpendicular to the field. 
(a) In the case of free electrons, the Landau level spacing is uniform with respect to the orbital quantum number $n$.
The averaged energy for $H\neq0$ is equal to that for $H\to0$, which is denoted by the horizontal dashed lines, so that the total energy for electrons deeply below the Fermi level is unchanged by introducing the field.
Only the highest level below the Fermi level contribute to the total energy change, which gives rise to the Landau diamagnetism.
(b) In the case of Dirac electrons, the energy level spacing is not uniform with respect to the total angular momentum $j$.
The averaged energy for $H\neq0$ is not equal to that for $H\to0$, so that the energy gain and energy loss are not cancelled for whole energy region. 
(The horizontal dashed lines are approximate values of the averaged energy for $H\to0$.
The blue (red) lines denotes the up (down) spins.)
Therefore, all energy levels below the Fermi level contribute to the increase of energy by magnetic field, i.e., diamagnetism, which is the largest when the Fermi level is in the gap.
}
\label{Landau_level}
\end{figure}
	
	We note here that the anomalous diamagnetism is already indicated by the form of the energy level of eq. (\ref{ene}), although detailed discussions will be given in the next section.

	For free electrons, the eigenvalues under a magnetic field is given by eq. (\ref{Estar}) with the bare electron mass.
	The orbital term and the Zeeman term are quantized independently, and give rise to the Landau diamagnetism and Pauli paramagnetism, respectively.
	By the orbital term, the electrons in the states
	\begin{align}
		\hbar \omega_{\rm c} n \leq \frac{\hbar^2 k_\perp^2}{2m_{\rm c}} \leq \hbar \omega_{\rm c} (n+1)
		\label{window}
\end{align}
	for the zero field limit ($H\to0$) shrinks into the level of $\hbar \omega_{\rm c}(n+1/2)$ as is shown in the left panel of Fig. \ref{Landau_level} (a).
	Since the density of states is uniform as a function of $k_\perp^2$ in the $\bk$-plane perpendicular to the field, the averaged energy in the interval (\ref{window}) for $H\to0$ (denoted by the horizontal dashed lines in Fig. \ref{Landau_level} (a)) is equal to $\hbar \omega_{\rm c} (n+1/2)$.
	Namely, the energy gain by introducing the magnetic field is equal to that of the energy loss in each energy interval. 
	Only the highest Lantau level below the Fermi level is related to the variation of the total energy.
	This gives rise to the Landau diamagnetism. 
	Similar cancellation happens also for the Zeeman term (the right panel of Fig. \ref{Landau_level} (a)), giving rise to the Pauli paramagnetism coming only from the vicinity of the Fermi level. 	

	For Dirac electrons, on the other hand, the orbital and the spin are completely mixed due to the strong spin-orbit interactions, so that the quantization is given only in terms of the total angular momentum $j$.
	The energy level splitting varies with respect to $j$, as is shown in Fig. \ref{Landau_level} (b).
	The averaged energy in the states
	\begin{align}
		\hbar \wc j \le \frac{\hbar^2 k_\perp^2}{ 2\mc} \le \hbar \wc (j+1)
\end{align}
	for $H\to 0$ does not agree with the averaged energy for $H\neq0$.
	The disagreement, which is shown by the shaded area in Fig. \ref{Landau_level} (b), exists in each energy interval, so that all energy levels below the Fermi level contribute the total energy.
	For electrons in the valence band, average energy for $H\neq0$ is always higher than the average energy for $H=0$ in each energy interval.
	The energy gain of the valence electrons by the magnetic field is always positive, resulting in the diamagnetism.
	The disagreement is the largest when the valence band is filled.
	This explains the fact that the diamagnetism becomes the largest when the Fermi level locates in the gap.
	%

\subsection{Anomalous magnetic-moment}
	
	Another important characteristics of Dirac electron in solids is the anomalous spin magnetic-moment.
	The electromagnetic properties of the Dirac electron are best exhibited by the transformation introduced by Foldy and Wouthuysen (FW)\cite{Foldy1950,Schweber_text}. 
	By their canonical transformation, the $4 \times 4$ Dirac Hamiltonian can be decoupled into the $2\times 2$ Hamiltonians for positive and negative energy.
	If we apply the FW transformation to the $4\times4$ Dirac Hamiltonian up to the order of $\hbar/mc$, we obtain two decoupled $2\times2$ Pauli Hamiltonians, $\scr{H}_{\rm P}=\pi^2/2m + \mu_{\rm B} \bsgm\cdot \bm{H}$.
	If we apply the FW transformation to the Wolff Hamiltonian up to the order of $|\bm{\gamma}| /\D$, we obtain
	\begin{align}
		\scr{H}_{\rm FW} &=
		\begin{pmatrix}
			\D + \scr{H}^* & 0\\
			0 & -\D -\scr{H}^*
		\end{pmatrix},
		\label{eq43}
\end{align}
	where $\scr{H}^*$ is the same as eq. (\ref{Hstar}).
	For the conduction band, the term linear in magnetic field is $(+\bm{\mu}_{\rm s}\cdot \bm{H})$, while, for the valence band, it is $(-\bm{\mu}_{\rm s}\cdot \bm{H})$.
	Therefore, the signs of $g^*$, eq. (\ref{gfactor}), are opposite between conduction and valence electrons.
	Correspondingly, the spin magnetic-moment of the Wolff Hamiltonian is defined in the form
		\begin{align}
		\bm{\mu}_{\rm s}=
		\begin{pmatrix}
			-\bm{\mu}_{\rm s}^* & 0 \\
			0 & \bm{\mu}_{\rm s}^*
		\end{pmatrix}.
		\label{magnetic moment}
\end{align}
	Since $g^*$ is quite large in bismuth\cite{Cohen1960,ZZhu2011}, the spin magnetic-moment is also quite large, $\mu_{\rm s}\sim 10^3 \mu_{\rm B}$.
	This large spin magnetic-moment can generate large spin responses such as in spin Hall effect and spin-polarized electric current.

\section{Diamagnetism}\label{Sec_Diamagnetism}

	The effect of a magnetic field in solids is very intricate due to the matrix elements between different Bloch bands (off-diagonal elements), referred to the interband effect, of a magnetic field. 
	As long as the off-diagonal terms are neglected, one can show that the energy of electrons under the magnetic field is given by changing the argument of the Bloch band as $E_n (\bk) \to E_n (\bk + (e/c)\bm{A})$.
	(Hereafter we take the unit of $\hbar=k_{\rm B}=1$.)
	When we apply this approximation to the orbital susceptibility, we obtain the Landau-Peierls formula\cite{Peierls1933,Peierls_text,Fukuyama1971}:
	\begin{align}
		\chi_{\rm LP}=\frac{e^2}{6\pi^3 c^2}\sum_{n,\bk}
		\left\{ \frac{\partial^2 E_n}{\partial k_x^2}\frac{\partial^2 E_n}{\partial k_y^2}
		-\left( \frac{\partial^2 E_n}{\partial k_x \partial k_y} \right)^2 \right\}
		\frac{\partial f(E_n)}{\partial E_n},
		\label{LP}
\end{align}
	where the magnetic field is applied along $z$-direction and $f(\ve)=[1+\exp\{ (\ve-\mu)/T\} ]^{-1}$ the Fermi distribution function.
	$\chi_{\rm LP}$ is proportional to $\partial f(E_n) /\partial E_n $, so that $\chi_{\rm LP}$ vanishes for insulators.
	As described in \S\ref{Sec_Intro}, the Landau-Peierls formula completely fails to explain the anomalously large diamagnetism of bismuth and its alloys (Fig. \ref{Wehrli}).
	After various efforts by many theorists\cite{Jones1934,Adams1953,Wilson1953,Kjeldaas1957,Hebborn1959,Hebborn1960,Roth1962,Blount1962,Brandt1963,Wannier1964,Ichimaru1965,Fukuyama1969,Fukuyama1970}, it was finally derived that the exact and gauge independent formula of diamagnetism can be given by a very simple one line formula as\cite{Fukuyama1970b,Fukuyama1971}
	\begin{align}
		\chi = \frac{ e^2}{2c^2}T\sum_{\sigma_z}\sum_{n, \bk}
		{\rm Tr} \left[ \scr{G}v_x \scr{G} v_y \scr{G} v_x \scr{G} v_x \right],
		\label{Fukuyama formula}
\end{align}
	where $n$ represents the index of Matsubara frequency, $\scr{G}$ is the thermal Green's function $\scr{G} (\bk , \Im \ve_n)$ of the matrix form, and $v_i$ is the velocity operator along the $i$-direction.
	Equation (\ref{Fukuyama formula}) was originally derived based on the Luttinger-Kohn representation, but can be considered to be that besed on the Bloch representation because of the trace-invariance of the representation.
	Surprisingly, this simple formula is proved to be equivalent to the previous exact but terribly complicated formulas\cite{Hebborn1959,Hebborn1960,Blount1962,Roth1962,Wannier1964,Ichimaru1965}.
	This formula (\ref{Fukuyama formula}) has been applied to the practical models of solids, e.g., graphene,\cite{Fukuyama2007} $\alpha$-ET$_2$I$_3$,\cite{Kobayashi2008} and bismuth.\cite{Fuseya2009,Fuseya2012b,Fuseya2014}

\subsection{Diamagnetism of Dirac electrons}
	%
	Here, we shall give the formulation for the orbital susceptibility of Dirac electrons in solids by the use of eq. (\ref{Fukuyama formula}). 
	We consider the Dirac Hamiltonian, eq. (\ref{isoWolff}), so that the Green's function, $\scr{G}=(\Im \tve_n - \scr{H})^{-1}$, and the velocity operator, $\bm{v}=\partial \scr{H}/ \partial \bk$, are given as
	\begin{align}
		\scr{G}(\bk, \Im \ve_n) 
		&=\frac{1}{(\Im \tve_n)^2 -E^2}
		\begin{pmatrix}
			\Im \tve_n + \D & \Im \gamma \bk \cdot \bsgm \\
			-\Im  \gamma \bk \cdot \bsgm & \Im \tve_n - \D
		\end{pmatrix},
		\\
		\bm{v}
		&=
		\begin{pmatrix}
			0 & \Im \gamma \bsgm\\
			-\Im \gamma \bsgm & 0
		\end{pmatrix},
\end{align}
	where $E^2 = \D^2 + \gamma^2 k^2$.
	We have introduced a finite damping, $\Gamma$, for electrons to represent the effects of impurity scattering as $\tve_n=\ve_n + \Gamma \sgn (\ve_n)$.
	We assume $\Gamma$ to be constant in order to make our argument as simple and transparent as possible, although Shon and Ando have indicated that $\Gamma$ somewhat depends on energy and momentum in case of graphene.\cite{Shon1998}

	The calculation of the trace of the eight $4\times 4$ matrices in eq. (\ref{Fukuyama formula}) is rather lengthy but straightforward.
	It becomes
	\begin{align}
		{\rm Tr}\left[ \scr{G}v_x \scr{G} v_y \scr{G} v_x \scr{G} v_y \right]
		&=-4\gamma^4 \frac{\{ (\Im \tve_n)^2 - E^2 \}^2 -8 \gamma^4 k_x^2 k_y^2 }{\{ (\Im \tve_n)^2 - E^2 \}^4}.
		\label{eq48}
\end{align}
	The summation and integration with respect to $\ve_n$ and $\bk$ results in
	\begin{align}
		\chi&=
		\frac{e^2 |\gamma|}{12\pi^2 c^2} \int_{-\infty}^{\infty}\!\! d\ve f(\ve)
		\left(\frac{1}{\sqrt{\ve_+^2 - \D^2}}-\frac{1}{\sqrt{\ve_-^2 - \D^2}}\right),
		\label{diamag_form}
\end{align}
	where $\ve_{\pm}=\ve\pm \Im \Gamma$.
	(Here the branch cut of the square root is taken along the positive real axis. The details of the calculation is given in Ref. \citen{Fuseya2014}.)
	The chemical potential dependence of $\chi$ is shown in Fig. \ref{results} (d).
	In the clean limit of $\Gamma\to 0$ and at zero temperature, we have
	\begin{align}
		\chi =
		\left\{
		\begin{array}{lc}
		\displaystyle
		-2\chi_0 \ln \left( \frac{E_{\rm c}}{|\mu|+\sqrt{\mu^2 -\D^2}} \right), & (|\mu|>\D) \\
		\displaystyle
		-2\chi_0 \ln \left( \frac{2E_{\rm c}}{\D}\right), & (|\mu|<\D)
		\end{array}
		\right. 
\end{align}
	where $\chi_0 = e^2 |\gamma| / 12\pi^2 c^2$ and $E_{\rm c}$ is the energy cutoff for the integration.
	The result is essentially equivalent to that obtained in Ref. \citen{Fukuyama1970}, which is derived based on the Wigner representation.

	As is shown in Fig. \ref{results} (d), $\chi$ takes the largest value for $|\mu|<\D$, all of which originates from the interband effect of the magnetic field.
	(The interband effect on $\chi$ appears more remarkably for the case of the two-dimensional Weyl Hamiltonian (valid for grahene) as shown in Fig. \ref{Fuku2007} (a).)

\section{Interband Effects on the Hall Effect}\label{Sec_Hall}
	In this section, we discuss the transport phenomena of Dirac electrons in solids.
	In order to keep the theoretical consistency with the results of the orbital susceptibility, we calculate the conductivities in the Luttinger-Kohn representation, and carefully analyze the interband effects resulting from a magnetic field.
	%

\subsection{Conductivities calculated by the Kubo formula}
	%
	The diagonal conductivity $\sigma_{xx}$ based on the Kubo formula\cite{Kubo1957,Fukuyama1969a,Fukuyama1969b} is given for the Dirac Hamiltonian in solids, eq. (\ref{isoWolff}), by
	\begin{align}
		\sigma_{xx} (\omega)
		&=\frac{e^2}{\Im \omega}
		\left[
		\Phi_{xx} (\omega + \Im \delta)
		-\Phi_{xx} (0 + \Im \delta)
		\right],
	\end{align}
		where
	\begin{align}
		\Phi_{xx}(\Im \omega_\lambda )
		&=-\frac{4e^2 \gamma^2}{\Im \omega}
		T \sum_n \sum_{\bm{k}}
		\frac{\Im \tilde{\ve}_n (\Im \tilde{\ve}_n-\Im \omega_\lambda )
		-\frac{1}{3}\gamma^2k^2-\Delta^2}
		{\left\{
		(\Im \tilde{\ve}_n)^2-E^2
		\right\}
		\left\{
		(\Im \tilde{\ve}_n - \Im \omega_\lambda)^2-E^2
		\right\}
		}.
	\end{align}
	Then we obtain the final expression in the form
		\begin{align}
		\sigma_{xx}
		&=
		-\frac{e^2}{\pi^3\gamma}
		\int_{-\infty}^\infty \!\!\!\! d\ve
		f' (\ve)
		\nonumber\\&\times
		\int_0^\infty \!\!\!\! dX
		\Biggl[
		\frac{X^2(\ve^2+\Gamma^2-\frac{1}{3}X^2-\Delta^2)}
		{\left( \ve_{+}^2-X^2-\Delta^2\right)
		\left( \ve_{-}^2-X^2-\Delta^2\right)}
		\nonumber\\
		&-\left\{
		\frac{X^2\left( \ve_{+}^2-\frac{1}{3}X^2-\Delta^2 
		\right)}
		{2\left(\ve_{+}^2-X^2-\Delta^2 \right)^2}
		+{\rm c.c.}
		\right\}
		\Biggr].
		\label{Cxx_eq}
	\end{align}

	The Hall conductivity $\sigma_{xy}$ based on the Kubo formula\cite{Fukuyama1969a,Fukuyama1969b,Fuseya2009} is also given by
	\begin{align}
		\sigma_{xy}&=
		\frac{1}{\Im \omega}
		\frac{8e^3 \gamma^4}{2c}(-\Im H) 
		T \sum_n \sum_{\bm{k}}
		\frac{\Im \omega_\lambda (2\Im \tilde{\ve}_n - \Im \omega_\lambda )}
		{\left\{ 
		(\Im \tilde{\ve}_n)^2 -E^2\right\}^2}
		\nonumber\\&\times
		\frac{\Im \tilde{\ve}_n (\Im \tilde{\ve}_n - \Im \omega_\lambda)
		+\gamma^2(k_x^2-k_y^2-k_z^2)-\Delta^2}
		{\left\{
		(\Im \tilde{\ve}_n -\Im \omega_\lambda)^2 -E^2
		\right\}^2}.
		\label{Cxy_eq}
	\end{align}
	With the same procedure as in $\sigma_{xx}$, we obtain
	\begin{align}
		\sigma_{xy}&=\frac{e^3 \gamma H}{12\pi^2c}
		\int_{-\infty}^\infty \!\!d\ve
		\left[
		K_{yx}^{\rm I}(\ve) f'(\ve)
		+
		K_{yx}^{\rm II}(\ve) f(\ve)
		\right]
		{\rm sgn}(\ve).
		\label{cxy}
		\\
		K_{yx}^{\rm I}(\ve)&=
		\frac{-2\Gamma^4-\Gamma^2\Delta^2+(\Delta^2-\ve^2)^2
		+2\Im \Gamma^3\ve -\Im \Gamma\ve(\Delta^2-\ve^2)}
		{2\Gamma^2\ve^2\sqrt{\ve_{+}^2-\Delta^2}}
		\nonumber\\&
		+{\rm c.c.}
		\\
		K_{yx}^{\rm II}(\ve)&=
		\frac{\ve_{+}}
		{( \ve_{+}^2-\Delta^2)^{3/2}}
		+{\rm c.c.}
\end{align}
	(Here, the branch cut of the square root is taken along the negative real axis.)
		%

\begin{figure}
\begin{center}
\includegraphics[width=8cm, bb=0 0 260 103]{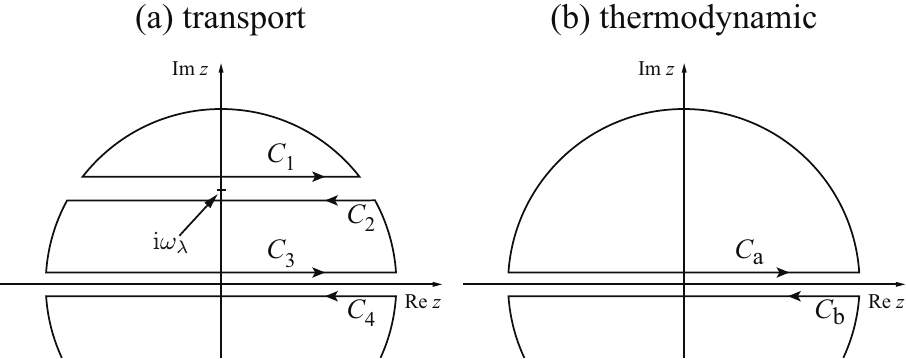}
\end{center}
\caption{Contours of integration in $z$-plane.
}
\label{f3}
\end{figure}
	%
	For the derivation of eqs. (\ref{Cxx_eq}) and (\ref{cxy}), we note the standard technique of the analytic continuation.
	We take the four contours $C_{1\sim 4}$ in the complex $z$ plane as shown in Fig. \ref{f3} (a).
	$C_1$ ($C_2$) is the contour from $-\infty$ to $\infty$ ($\infty$ to $-\infty$) along just above (below) the horizontal line ${\rm Im}\, z = \omega_\lambda$.
	$C_3$ ($C_4$) is the contour from $-\infty$ to $\infty$ ($\infty$ to $-\infty$) along just above (below) the horizontal line ${\rm Im}\, z = 0$.
	Basically, the $f'$-term (e.g., the first term of eq. (\ref{cxy})) originates from the contribution of $C_2 + C_3$, which has a functional form of
	\begin{align}
		-\frac{1}{2\pi \Im}\int_{-\infty}^{\infty}\!\! d\ve f(\ve)
		\left[ G^{\rm R}(\ve)G^{\rm A}(\ve-\omega) - G^{\rm R}(\ve+\omega)G^{\rm A}(\ve) \right],
		\label{GRGA}
\end{align}
	where $G^{\rm R (A)}$ is the retarded (advanced) Green's function.
	We only need the $\omega$-linear term for the static response, and it is obtained by shifting the variable $\ve \to \ve-\omega$ in the second term of eq. (\ref{GRGA}) as
	\begin{align}
		\frac{\Im \omega}{2\pi} \int_{-\infty}^{\infty}\!\! d\ve 
		\left( \frac{\partial f(\ve)}{\partial \ve} \right)G^{\rm R}(\ve)G^{\rm A}(\ve).
\end{align}
	It is proportional to $f'(\ve)=\partial f(\ve)/\partial \ve$, so that the states only in the vicinity of the Fermi level contribute to this term.

	The $f$-term (e.g., the second term of eq. (\ref{cxy})) originates from the contribution from the contours $C_1 + C_4$, which has a functional form of
	\begin{align}
		\omega f(\ve) \left[ G^{\rm R}(\ve) G^{\rm R}(\ve) - G^{\rm A}(\ve) G^{\rm A}(\ve) \right],
\end{align}
	for the term linear in $\omega$.
	This term is similar to the contribution in the thermodynamical quantities. 
	This similarity can be understood if we consider the way to take the limit.
	For the transport quantities, we first take the $\bq \to 0$ limit, and then take the $\omega \to 0$ limit, so that we have to consider the four contours $C_1 \sim C_4$ due to the finite $\omega_\lambda$.
	For the thermodynamical quantities, on the other hand, we first take the $\omega \to 0$ limit, and then take the $\bq \to 0$ limit, so that we need only the two contours $C_a$ and $C_b$ in Fig. \ref{f3} (b).
	We can clearly see from Fig. \ref{f3} that contribution from $C_1 + C_4 $ has the same analytic property as $C_a + C_b$.
	The relationship between the contribution from $C_1 + C_4$ and that from $C_a + C_b$ will be discussed again for the relationship between the spin Hall effect and the diamagnetism in \S \ref{Sec_SHE}.

	Note that, in eq. (\ref{Cxx_eq}), the contribution from $C_1 + C_4$, which is originally proportional to $f(\ve)$, is rewritten only by the term that is proportional to $f'(\ve)$ using the integration by parts in $\sigma_{xx}$.
	Thus in eq. (\ref{Cxx_eq}), the total contribution is expressed as $f'$-term, though it includes both $C_2 + C_3$ and $C_1 + C_4$ contributions.
	On the other hand, $\sigma_{xy}$ cannot be expressed only in terms of $f'$-term, even if we use the integration by parts.

\begin{figure}
\begin{center}
\includegraphics[width=70mm, bb=0 0 549 1325]{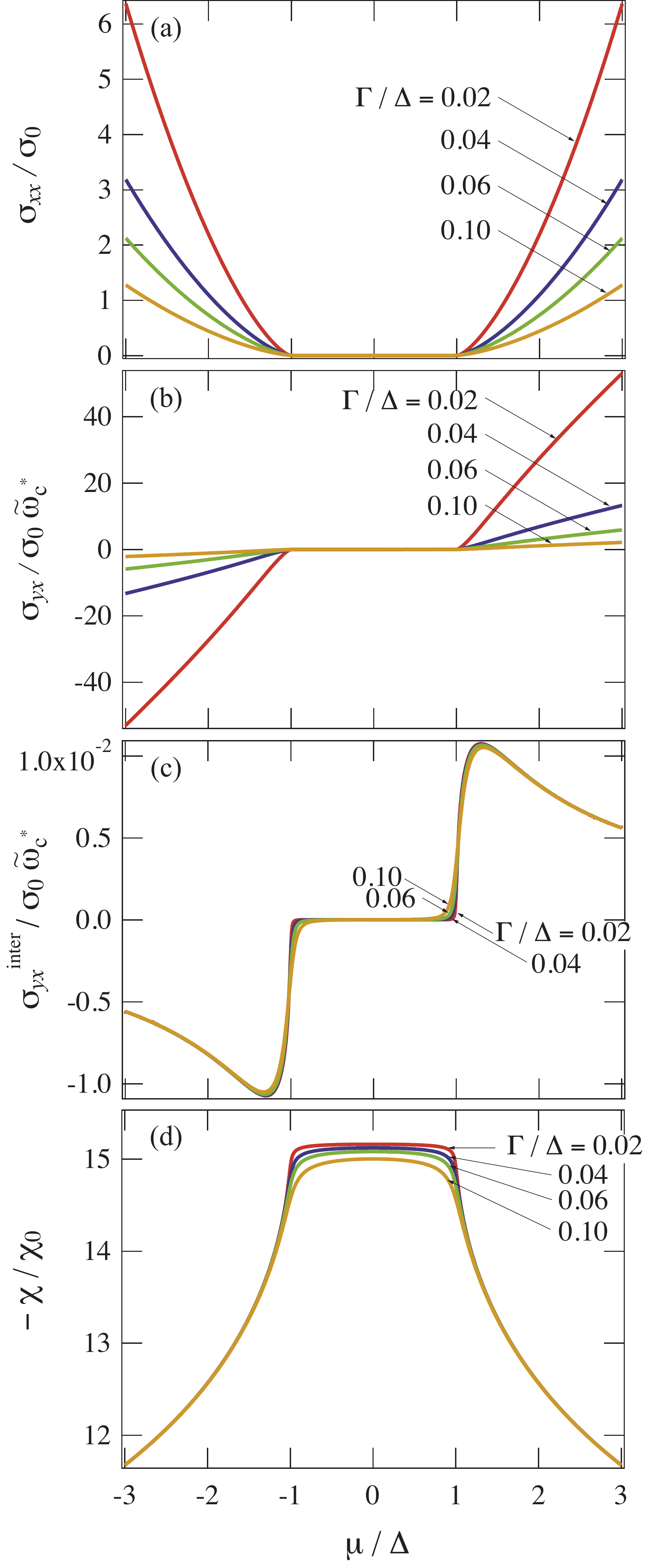}
\end{center}
\caption{(Color online) Theoretical results of the chemical potential dependences for (a) the conductivity $\sigma_{xx}$, (b) the Hall conductivity $\sigma_{yx}$, (c) the interband contribution to $\sigma_{yx}$, and (d) the orbital susceptibility. 
For normalization, we used $\sigma_0 = e^2 \D/\gamma$, $\tilde{\omega}_{\rm c}^* = \wc / \D$ and $\chi_0 = e^2 |\gamma|/12\pi^2 c^2$.
}
\label{results}
\end{figure}
	%
	
	The chemical potential dependence of $\sigma_{xx}$ and $\sigma_{yx}$ at zero temperature are shown in Fig. \ref{results} (a) and (b), respectively.
	In the case of bismuth, we can control $\mu$ by substituting other elements, e.g., Sn, Pb for hole doping, and Te, Se for electron doping.
	In the metallic region,  $|\mu| \gg \Delta$, $\sigma_{xx}\propto\mu^2$ and $\sigma_{yx}\propto \mu$.
	In the insulating region, $|\mu|<\Delta$, both conductivities are vanishingly small.
	(Strictly speaking, they take small values, which is due to the finite $\Gamma$.)
	Our results (Fig. \ref{results} (a) and (b)) seem to agree quite well with the results obtained by the Boltzmann (intraband) approximation, $\sigma_{xx}\propto \mu^{2}$ and $\sigma_{xy}\propto \mu$, at first sight.
	Also, the results exhibit the strong $\Gamma$-dependence both in $\sigma_{xx}$ and $\sigma_{xy}$, which are quite natural as the transport phenomena.
	However, interesting properties are found, if we look carefully into the interband effect on $\sigma_{xy}$.

\subsection{Interband effects on the Hall conductivity}\label{InterSxy}

	Now we look into the interband effects of a magnetic field on $\sigma_{xy}$ in detail.
	The Hall conductivity formula obtained by eq. (\ref{cxy}) is exact, which includes both intra- and inter-band contributions.
	The interband contribution, $\sigma_{xy}^{\rm inter}$, can be defined by subtracting the intraband contribution, $\sigma_{xy}^{\rm intra}$, from eq. (\ref{cxy}):
\begin{align}
	\sigma_{xy}^{\rm inter}= \sigma_{xy}-\sigma_{xy}^{\rm intra}.
	\label{def_inter}
\end{align}

	The intraband contribution is calculated by the Bloch band picture, which is usually given by ($\Gamma = 1/2\tau$)
	\begin{align}
	\sigma_{xy}^{\rm Boltzmann}&=-\frac{2e^3 H \tau^2}{c}\sum_{n,\bk}
	\left\{
		\frac{\partial^2 E_n}{\partial k_x^2}\frac{\partial^2 E_n}{\partial k_y^2}
		-\left( \frac{\partial^2 E_n}{\partial k_x \partial k_y} \right)^2
	\right\} \frac{\partial f(E_n)}{\partial E_n},
	\label{Boltzmann}
	\end{align}
	when the magnetic field is along $z$-direction.
	In the present case of Dirac Hamiltonian in solids, we have\cite{Fuseya2009}
	\begin{align}
	\sigma_{xy}^{\rm intra}
	&=-\frac{e^3vH}{6\pi^3c}\sum_{n=\pm}
	\int_{-\infty}^{\infty}\!\! d\ve 
	f'(\ve)
	\int_0^\infty \!\! dX
	\frac{X^4 }{\left[ E_n (X) \right]^3}
	\nonumber\\
	&\times
	\frac{4\Gamma^3}{3\left[ (\ve-E_n(X))^2+\Gamma^2\right]^3},
	\label{cxy_intra}
\end{align}
	where $E_\pm(X)=\pm\sqrt{X^2+\Delta^2}$.
	The expression of eq. (\ref{cxy_intra}) is quite different from that of Eq. (\ref{cxy}).
	Nevertheless, they agree with each other {\it except for the band-edge region}, showing the validity of the Bloch band approximation for the metallic region.
	Near the band-edge region, on the other hand, the interband contribution $\sigma_{xy}^{\rm inter}$ is remarkable.

	The obtained $\sigma_{xy}^{\rm inter}$ is shown in Fig. \ref{results} (c).
	The properties of $\sigma_{xy}^{\rm inter}$ are found to be as follows:
	\begin{description}
		\item[{\rm (1)}] $\sigma_{xy}^{\rm inter}$ takes the {\it maximum} value near the band-edge,
		
		\item[{\rm (2)}]  {\it decreases} away from the band-edge as $\sigma_{xy}^{\rm inter} \propto -\mu^{-1}$,
		
		\item[{\rm (3)}] {\it weakly} depends on $\Gamma$,
		
		\item[{\rm (4)}] magnitude of $\sigma_{xy}^{\rm inter}$ is much smaller than that of $\sigma_{xy}^{\rm intra}$.
	\end{description}
	The properties (1)-(3) are completely different from that of $\sigma_{xy}^{\rm intra}$.
	(The properties of $\sigma_{xy}^{\rm intra}$ is roughly the same as $\sigma_{xy}$ shown in Fig. \ref{results} (b), since the $\sigma_{xy}^{\rm inter}$ is much smaller than $\sigma_{xy}^{\rm intra}$.)
	These differences reveal that $\sigma_{xy}^{\rm inter}$ is not generated by the normal conduction electrons.
	By contrast, the properties of $\sigma_{xy}^{\rm inter}$ is quite similar to the orbital susceptibility shown in Fig. \ref{results} (e).
	These similarities strongly suggest that $\sigma_{xy}^{\rm inter}$ has a common origin with the diamagnetic current.
	%
	
\begin{figure}[tb]
\begin{center}
\includegraphics[width=8cm, bb=0 0 603 223]{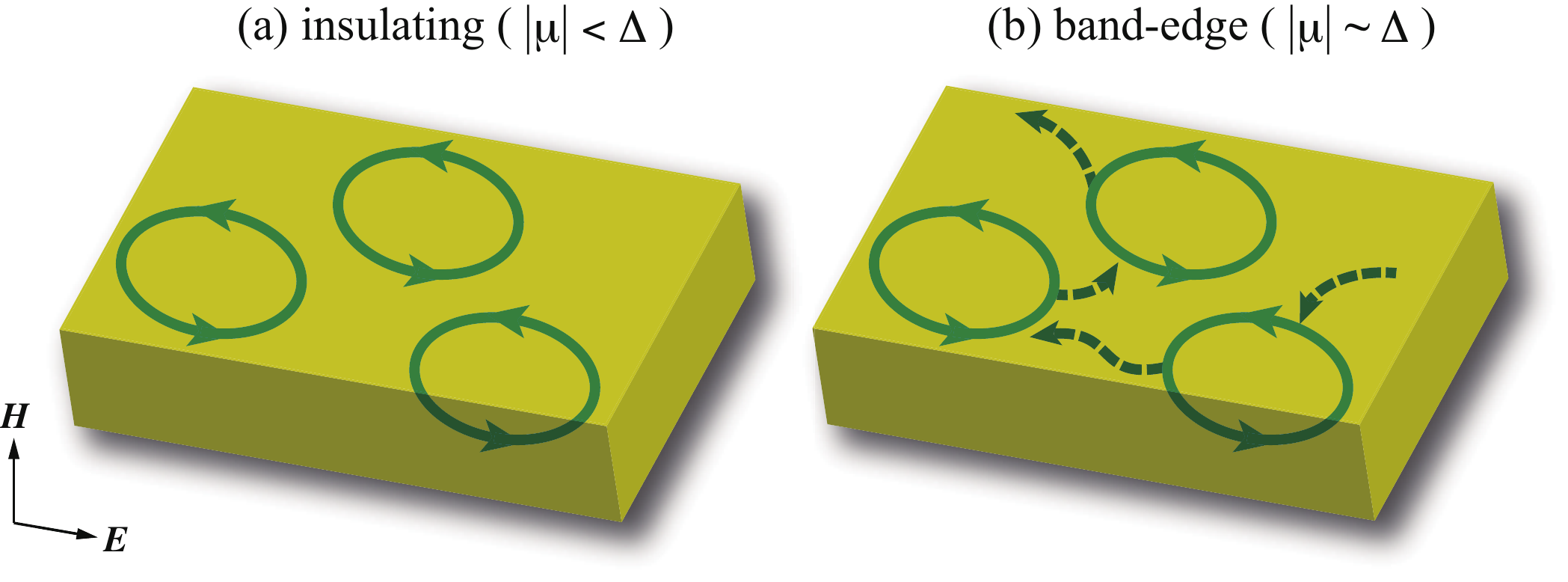}
\end{center}
\caption{\label{orbital}
(Color online) Schematic motion of electrons in a solid for (a) the insulating ($|\mu|<\Delta$), (b) the band-edge ($|\mu|\sim \Delta$). 
}
\end{figure}
	%
	Their relationship is interpreted as follows.
	In the insulating region, electrons in a magnetic field make a local orbital motion, and generate the diamagnetic current, which is persistent and dissipationless (Fig. \ref{orbital} (a)).
	%
	%
	There are no electrons going through the crystal, so that they do not contribute to the Hall conductivity.
	%
	%
	In the band-edge region, the electrons belong to the local diamagnetic orbital start to transfer to the neighboring orbitals via hybridization with the conduction electrons (Fig. \ref{orbital} (b)).
	This time electrons go through the crystal by this transfer between local diamagnetic orbitals and can contribute to the Hall conductivity as $\sigma_{xy}^{\rm inter}$.
	The local diamagnetic current has largest values for $|\mu|\le \Delta$ and decreases away from the band-edge.
	Correspondingly, $|\sigma_{xy}^{\rm inter}|$ actually decreases away from the band-edge.
	With this picture, $\sigma_{xy}^{\rm inter}$ slightly contains the conduction electron current, but the main contribution comes from the diamagnetic current, so that the dissipation of $\sigma_{xy}^{\rm inter}$ will be vanishingly small.
	%
	
\begin{figure}
\begin{center}
\includegraphics[width=70mm, bb=0 0 537 387]{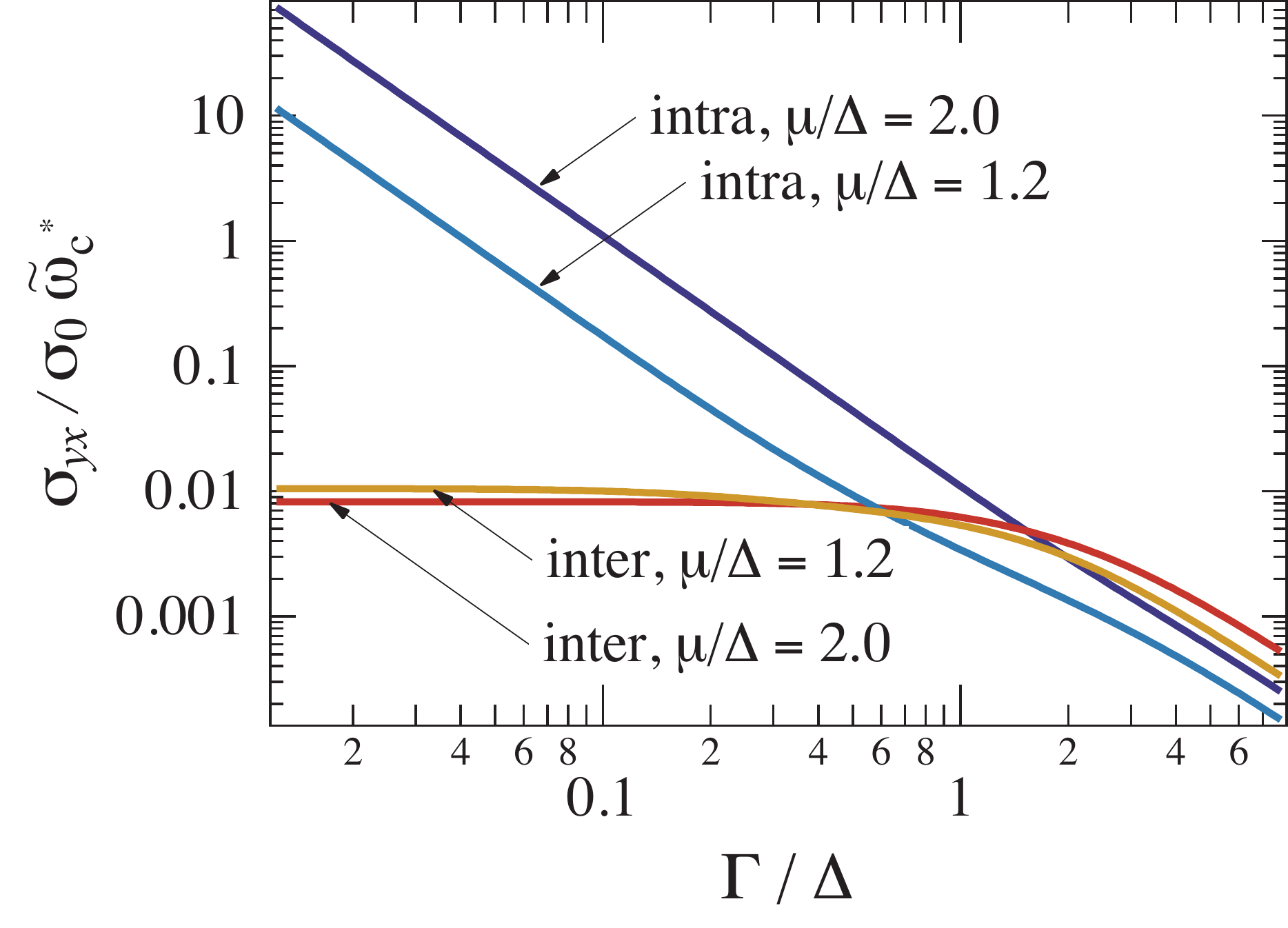}
\end{center}
\caption{(Color online) Impurity scattering dependence of $\sigma_{yx}^{\rm intra, inter}$ for $\mu/\D =1.2$ (band-edge), and 2.0 (metallic).
}
\label{cxy_inter_gamma}
\end{figure}
	%
	
	%
	The $\Gamma$-dependence of $\sigma_{xy}$ is shown in Fig. \ref{cxy_inter_gamma}.
	For the band-edge ($\mu=1.2$) and the metallic ($\mu=2.0$) region, $\sigma_{xy}^{\rm inter}$ does not depend on $\Gamma$, while $\sigma_{xy}^{\rm intra} \propto \Gamma^{-2}$ for $\Gamma < \D$.
	This means that $\sigma_{xy}^{\rm inter}$ is not affected by the impurity scatterings, suggesting that the dissipation of $\sigma_{xy}^{\rm inter}$ is much smaller than the ordinal current.
	From these facts, we can conclude that we would have a current that hardly generate Joule heat, if we could have $\sigma_{xy}^{\rm inter}$ only.
	Unfortunately, however, $\sigma_{xy}^{\rm inter}$ always flows with $\sigma_{xy}^{\rm intra}$, and $\sigma_{xy}^{\rm inter} \ll \sigma_{xy}^{\rm intra}$, so that we cannot obtain $\sigma_{xy}^{\rm inter}$ only at least for dc current.
	On the contrary, in the ac conductivity, we can take out $\sigma_{xy}^{\rm inter}$ selectively by tuning a frequency.
	This will be discussed in \S\ref{Sec_SPEC}.
	%

\subsection{Hall coefficient near the band-edge}
\begin{figure}
\begin{center}
\includegraphics[width=70mm, bb=0 0 523 387]{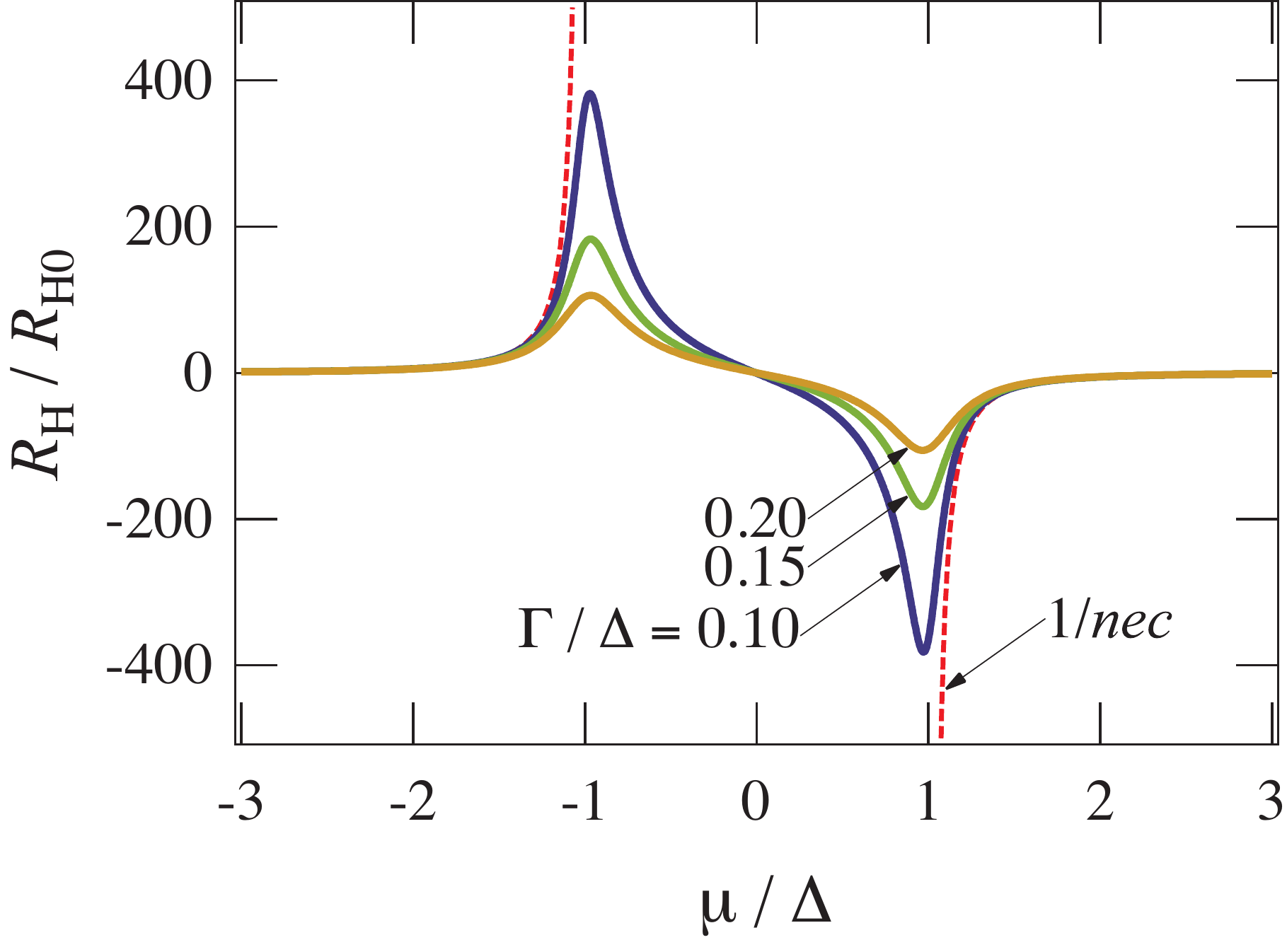}
\end{center}
\caption{(Color online) Chemical potential dependences of the Hall coefficient, $R_{\rm H}=\sigma_{xy}/\sigma_{xx}^2H$, for the Dirac electrons in solids. The chemical potential dependences of $(1/nec)$ is also shown by the dashed line.
For normalization, we used $R_{\rm H0}=\tilde{\omega}_{\rm c}^*/\sigma_0 H$.
}
\label{RH}
\end{figure}
\begin{figure}
\begin{center}
\includegraphics[width=50mm, bb=0 0 155 348]{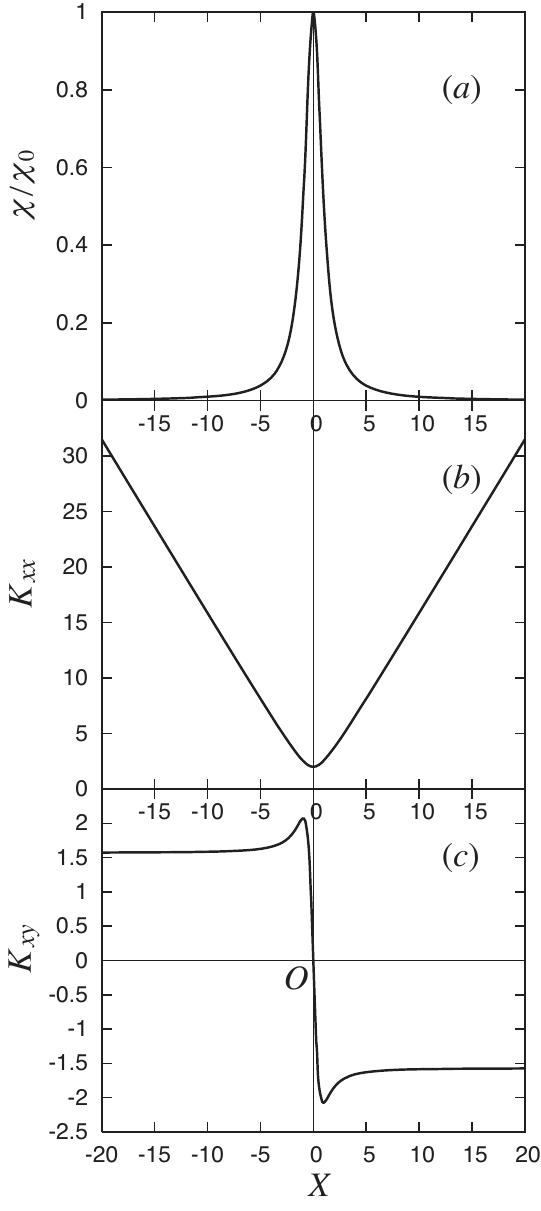}
\end{center}
\caption{Theoretical results of (a) orbital diamagnetism, (b) the conductivity and (c) the Hall conductivity in two-dimensional Weyl Hamiltonian (valid for graphene) as a function of $X=\mu/\Gamma$. 
Taken from Ref. \citen{Fukuyama2007}
}
\label{Fuku2007}
\end{figure}
\begin{figure}
\begin{center}
\includegraphics[width=70mm, bb=0 0 243 189]{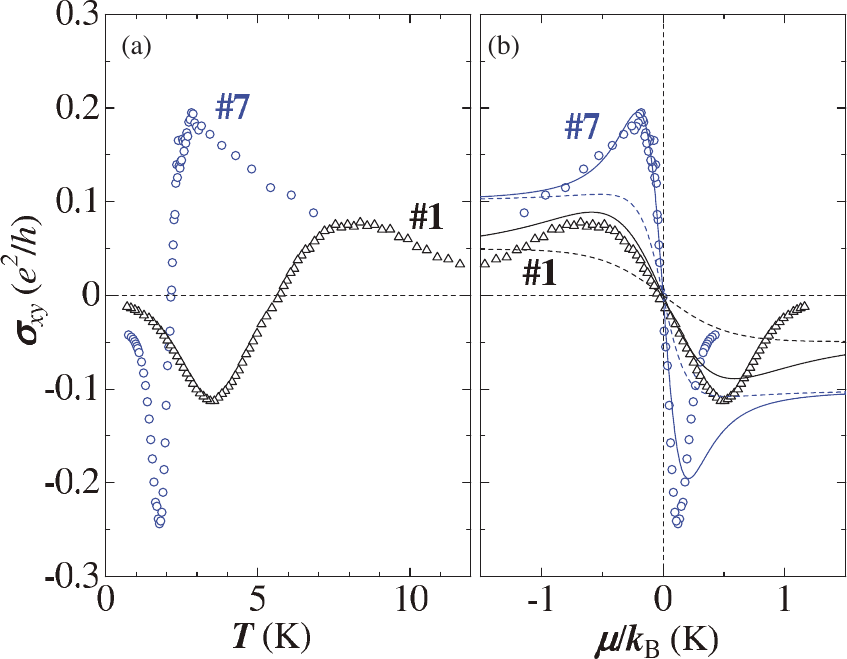}
\end{center}
\caption{(Color online) (a) Temperature dependence of $\sigma_{xy}$ for different samples (denoted as \#1 and \#7).
(b) Chemical potential dependence of $\sigma_{xy}$ for sample \#1 and \#7.
Solid lines and dashed lines are the theoretical results with and without the interband contributions by Kobayashi {\it et al.}.\cite{Kobayashi2008}
Taken from Ref. \citen{Tajima2012}.
}
\label{Tajima2012}
\end{figure}
	%
	We also give another non-trivial property that appears near the band-edge region.
	The Hall coefficient defined by $\RH=\sigma_{xy}/\sigma_{xx}^2 H$ with eqs. (\ref{Cxx_eq}) and (\ref{Cxy_eq}) is shown in Fig. \ref{RH}. 
	$\RH$ gives a good measure of carrier density in the metallic region.
	However, at the band-edge $|\mu| \simeq \D$, $\RH$ strongly deviates from $1/nec$ implying that one can not estimate carrier density from $\RH$.
	Especially, $\RH$ is vanishingly small and changes its sign at $\mu=0$ indicating that ``the effective carrier density" is diverging.
	Therefore, $\RH$ is nothing to do with $n$ for $|\mu|\lesssim \D$.

	These sign reversal in $\RH$ or $\sigma_{xy}$ appears dramatically in the zero-gap case, such as the graphene and $\alpha$-ET$_2$I$_3$; $\sigma_{xy}$ changes the sign when the chemical potential varies through the crossing points (Fig. \ref{Fuku2007} (c)).
	This sign reversal of $\sigma_{xy}$ is actually observed in $\alpha$-ET$_2$I$_3$\cite{Tajima2012,Kajita2014,note8} (Fig. \ref{Tajima2012}).
	%

\section{Spin Hall Effect}\label{Sec_SHE}

	In the previous section, we saw there is a certain relationship between the diamagnetism and the interband Hall conductivity.
	In addition to this, it has been found recently that there is a surprising relationship between the spin Hall effect (SHE) and diamagnetism.\cite{Fuseya2012b,Fuseya2014}

\begin{figure}
\begin{center}
\includegraphics[width=6cm, bb=0 0 304 197]{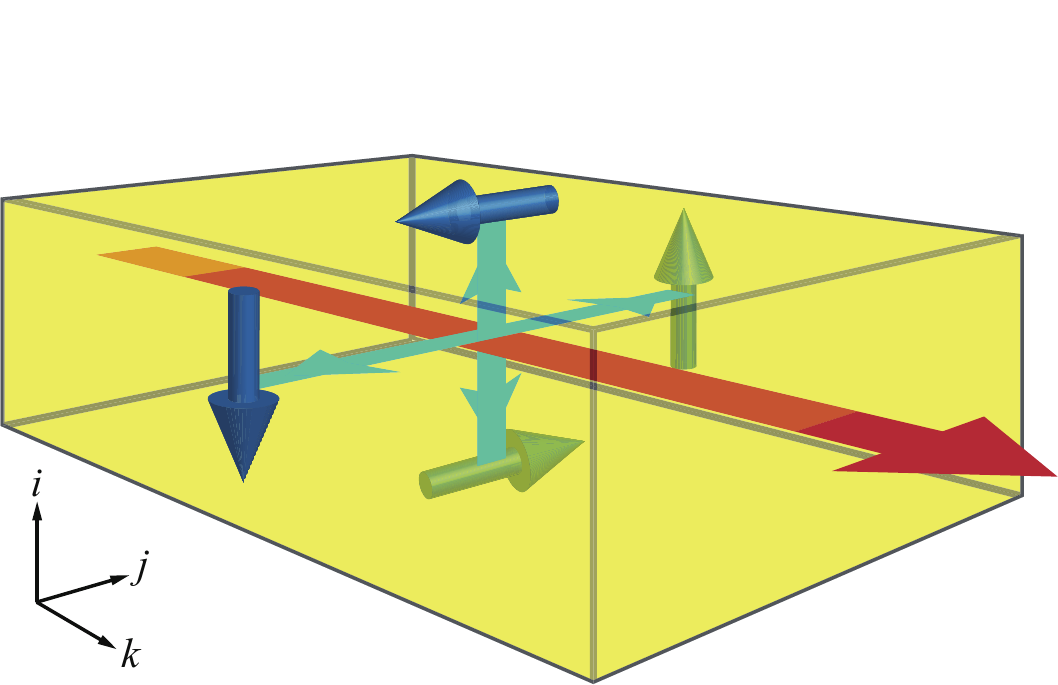}
\end{center}
\caption{(Color online) 
Illustration of the directions of the electric field (wide arrow), the spin current (narrow arrows) and the spin magnetic-moment (three-dimensional arrows).
}
\label{Illust_SHE}
\end{figure}

	The SHE is an effect that the spin magnetic-moment of electrons experience a transverse force in the presence of a longitudinal electric field\cite{Hirsch1999} (Fig. \ref{Illust_SHE}).
	The essential idea was first introduced by Dyakonov and Perel\cite{Dyakonov1971a,Dyakonov1971b}:
	a spatial separation of electrons with different spins is generated due to the spin-orbit interaction via the scattering of unpolarized electrons by an unpolarized target.
	This can lead to spin-accumulation at the edge of a thin layer.
	The idea was introduced as a counterpart of the anomalous Hall effect. 
	%
	
	A renaissance of interest in this subject was revived by Hirsch, who named this effect ``spin Hall effect".\cite{Hirsch1999}
	He proposed the practical way for the experimental detection of SHE.
	Soon later, Zhang extended the idea of SHE to the diffusive transport regime, and suggested another experiment\cite{SZhang2000}.
	These proposals activated the research in this field.
	The SHE was observed by
	the magneto-optical Kerr effect measurements,\cite{Kato2004,Sih2005,Sih2006,Stern2006}
	the circularly polarized electroluminescence,\cite{Wunderlich2005}
	and the non-local electrical measurement in nanoscale H-shaped structure.\cite{Brune2010}
	Also, the inverse SHE, where polarized electron current generates a transverse voltage, was observed by injecting a spin current from a ferromagnetic electrode into a non-magnetic metal.\cite{Saitoh2006,Valenzuela2006}
	It was experimentally confirmed that both the direct and inverse SHE are equivalent, demonstrating the Onsager reciprocal relations.\cite{Kimura2007}
	%


	The suggested mechanisms by Dyakonov-Perel and Hirsch are basically the effect driven by the scattering due to impurities, the so-called  extrinsic mechanism.
	On the other hand, the possibility of an intrinsic mechanism of SHE was put forward by Murakami {\it et al.}\cite{Murakami2003} and Sinova {\it et al}\cite{Sinova2004}. 
	Unlike the extrinsic mechanism, where the scattering plays a dominant role, the intrinsic SHE is the effect which originates purely from the electronic structure of solids; the scattering contributions are less dominant.
	It is closely related to the intrinsic mechanism of the anomalous Hall effect.
	The intrinsic SHE is possible not only for metals, but also for semiconductors or insulators. 
	%
	%
	These proposals of intrinsic SHE triggered intensive studies of theory and experiment, which is summarized in several review articles.\cite{Murakami2006b,Sinova2006,Engel2007,Sinova2008,Vignale2010,Murakami2011}
	In this section, we discuss the intrinsic SHE for Dirac electrons.

\subsection{SHE of Dirac electrons}
	It is naively expected that the materials composed of elements with a large spin-orbit interaction will exhibit a large SHE.
	In this context, bismuth will be one of the best candidates since the spin-orbit interaction of bismuth atom is the largest ($\sim 1.5$ eV) among the non-radioactive elements.\cite{note3}
	However, the effect of the spin-orbit interaction in crystals does not directly correspond to the magnitude of the spin-orbit interaction of isolated atoms.
	 It is nontrivial whether the crystalline bismuth exhibits a large SHE or not.
	Here we show the results of SHE on the crystalline bismuth.

	We consider the simplest situation: apply a longitudinal electric field, and measure the difference in magnetization between the edges perpendicular to the field, e.g. by the magneto-optical Kerr effect\cite{Kato2004,Sih2005,Sih2006,Stern2006}.
	%
	%
	The difference of magnetization is caused by the imbalance of the spin magnetic-moment perpendicular to the field, which is generated by the flow of the spin magnetic-moment.
	Based on this consideration, we calculate the current of the spin magnetic-moment, $j_i^z = \left\{ \mu_{{\rm s}z}, v_i \right\}/2$.
	Although this is basically the same as the definition $j_i^z = \left\{s_z, v_i \right\}/2$ for the spin current, we stress that we pay attention not to the spin, but to the spin magnetic-moment, since the observable physical quantity is the magnetization.
	The definition of spin current and the relationship to the continuation equation are discussed in detail in Appendix.
	
	We start from the Dirac Hamiltonian of eq. (\ref{isoWolff}).\cite{Fuseya2012b}
	The spin magnetic-moment of this Hamiltonian is given by (cf. eq. (\ref{magnetic moment}))
	\begin{align}
		\bm{\mu}_{\rm s}=\frac{g^* \mu_{\rm B}}{2}
		\begin{pmatrix}
			-\bsgm & 0 \\
			0 & \bsgm
		\end{pmatrix},
\end{align}
	 where the sign of the magnetic moment is opposite between the conduction and valence band.
	 (The g-factor is given by eq. (\ref{gfactor}) with $\mc=\D/\gamma^2$ for the present Hamiltonian.)
	Correspondingly, we introduce the velocity operator of the spin magnetic-moment as
	\begin{align}
		v_{{\rm s}i}= \frac{\mu_{{\rm s}z}v_i}{\mu_{\rm B}}=-\Im \frac{g^* \gamma}{2}
		\begin{pmatrix}
			0 & \sigma_z \sigma_i \\
			\sigma_z \sigma_i & 0
		\end{pmatrix},
		\label{spin velocity}
\end{align}
	where $i=x, y$.
	(We normalized by $\mu_{\rm B}$ in order to fit its dimension to the ordinary dimension of the velocity.)
	In the present case, this is an Hermitian operator.
	Hereafter we call this velocity operator of the spin magnetic-moment as the spin current.
	
	The spin Hall conductivity (SHC) is given as a linear response of the transverse spin current to the electric field on the basis of the Kubo formula:
	\begin{align}
		\sigma_{{\rm s}yx}&=\frac{1}{\Im \omega}
		\left[ \Phi_{{\rm s}yx}(\omega + \Im \delta) - \Phi_{{\rm s}yx}(0+\Im \delta ) \right],
		\label{eq91}
		\\
		\Phi_{{\rm s}yx}(\Im \omega_\lambda) &= -eT
		\sum_{n \bk} {\rm Tr}
		\left[ \scr{G}(\Im \tve_n)v_{{\rm s}y}\scr{G}(\Im \tve_{n-})v_x \right]
		\label{eqpre65}\\
		&=
		eT \sum_{n \bk} \frac{4\Im m\gamma^4 (\Im \tve_n - \Im \tve_{n-})}
		{\{(\Im \tve_n )^2 - E^2\}\{(\Im \tve_{n-})^2 - E^2\}}.
		\label{eq65}
\end{align}
	where $\ve_{n-}=\ve_n - \omega_\lambda$.
	After the $n$-summation by the analytic continuation (see \S\ref{Sec_Hall}), and the $\bk$-integration, we obtain
	\begin{align}
		\sigma_{{\rm s}yx}&=-\frac{em|\gamma|}{4\pi^2}
		(K_{{\rm s}yx}^{\rm I} + K_{{\rm s}yx}^{\rm II}),
		\label{K0}
		\\
		K_{{\rm s}yx}^{\rm I}&= \int_{-\infty}^{\infty} \!\! d\ve
		\left( \frac{\partial f(\ve)}{\partial \ve} \right)
		\left[
		\frac{\sqrt{\ve_+^2 - \D^2}}{\ve} - \frac{\sqrt{\ve_-^2 - \D^2}}{\ve}
		\right],
		\label{KI}
		\\
		K_{{\rm s}yx}^{\rm II}&= \int_{-\infty}^{\infty} \!\! d\ve
		f(\ve)
		\left[
		\frac{1}{\sqrt{\ve_+^2 -\D^2}}-\frac{1}{\sqrt{\ve_+^2 -\D^2}},
		\right]
		\label{KII}
\end{align}
	where $\ve_{\pm}=\ve \pm \Im \Gamma$. 
	(Here the branch cut of the square root is taken along the positive real axis.)
	Similarly, we can calculate the diagonal spin conductivity $\sigma_{{\rm s}xx}$, but it is exactly zero.
	For the clean limit, $\Gamma / \D \to 0$, at zero temperature, eqs. (\ref{KI}) and (\ref{KII}) are given by the following simple forms:
	\begin{align}
		-K_{{\rm s}yx}^{\rm I} &=
		\left\{
		\begin{array}{cc}
		\displaystyle \frac{2\sqrt{\mu^2 -\D^2}}{|\mu|} & \left(|\mu|>\D \right) \\
		\\
		0 & \left(|\mu|<\D\right)
		\end{array}
		\right.,
		\label{KIclean}
		\\
		-K_{{\rm s}yx}^{\rm II} &=
		\left\{
		\begin{array}{cc}
			\displaystyle 2\ln \left|\frac{2E_{\rm c}}{|\mu| + \sqrt{\mu^2-\D^2}}\right| & \left(|\mu|>\D \right) \\
			\\
			\displaystyle 2 \ln \left| \frac{2E_{\rm c}}{\D}\right| & \left(|\mu|<\D\right)
		\end{array}
		\right.,
		\label{KIIclean}
\end{align}
	where $E_{\rm c}$ is the energy cutoff for the integration, and we discarded $\mathcal{O}(\D^2 /E_{\rm c}^2)$-term.

\begin{figure}
\begin{center}
\includegraphics[width=7cm, bb=0 0 487 388]{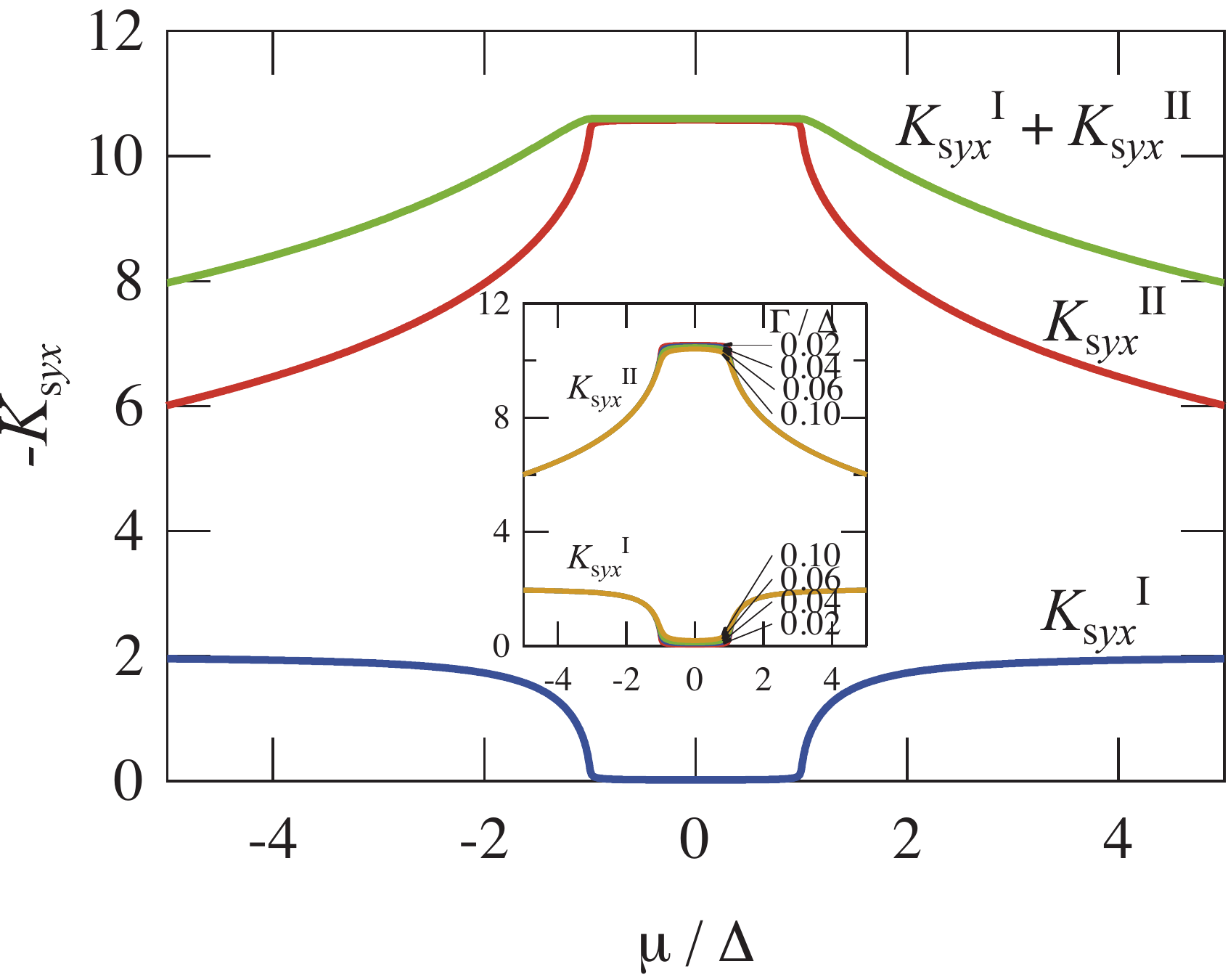}
\caption{(Color online) Chemical potential dependence of the SHC ($K_{{\rm s}yx}^{\rm I, II}$) at zero temperature for $\Gamma/\D=0.01$ and $E_{\rm c}/\D=100$.
The inset shows the plot of $K_{{\rm s}yx}^{\rm I, II}$ for different ramping rates: $\Gamma \D = 0.02, 0.04, 0.06, 0.10$.}
\label{SHE_mu}
\end{center}
\end{figure}
	
	%
	The chemical-potential dependences of $K_{{\rm s}yx}^{\rm I, II}$ are shown in Fig. \ref{SHE_mu}.
	We see that the SHE takes maximum in the insulating region, and it decreases as the carrier density increases;
	this result recalls the property of the diamagnetism of Dirac electrons.
	Actually, the form of $K_{{\rm s}yx}^{\rm II}$, eq. (\ref{KII}) is exactly the same as that of orbital susceptibility, eq. (\ref{diamag_form}), except for their coefficients.
	Their starting points, eqs. (\ref{eq48}) and (\ref{eq65}), are completely different, but their final results are equivalent --- an amazing fact.
	Therefore, we obtain a very simple formula that relates the SHC to the orbital susceptibility in the insulating region:
	\begin{align}
		\sigma_{{\rm s}xy}=\frac{3mc^2}{e} \chi.
		\label{SHC_chi}
\end{align}
	In the insulating region, $\sigma_{xx}$ is suppressed, so that there are no dissipative current.
	Only in such a case, the SHE is dissipationless and becomes exactly parallel to the diamagnetic susceptibility. 
	The SHE is generated by the external electric field, while the diamagnetism is by the external magnetic field.
	They are completely different phenomena, but are related with each other through eq. (\ref{SHC_chi}).

	The relationship between the SHC and the spin density has been argued in the context of the two-dimensional quantum SHE.\cite{Yang2006}
	They considered ``spin conserved" part of the SHC in the insulating case is given by a St\v{r}eda-like formula, $\sigma_{{\rm s}yx}^{\rm II, (c)} =-\partial S_z / \partial H$, where $S_z$ is the $z$-component of the spin density.
	This is consistent with eq. (\ref{SHC_chi}).
	However, $S_z$ is not an observable physical quantity, so that their relation in terms of $S_z$ cannot be checked experimentally.
	In contrast, eq. (\ref{SHC_chi}) connects the SHC with the observable magnetic susceptibility.
	We also note that the St\v{r}eda-like formula used in Ref. \citen{Yang2006} is valid only for the insulating case, while the calculation based on the Kubo formula, eq. (\ref{eq91}), is valid both for insulating and metallic case.


	%
	We can divide $\Phi_{{\rm s}yx}(\Im \omega_\lambda)$ into the contributions from the intra- and inter-band effects as follows.
	\begin{align}
		\Phi_{{\rm s}yx}(\Im \omega_\lambda)&=\sum_{\alpha \beta} \Phi_{{\rm s}yx}^{\alpha \beta}(\Im \omega_\lambda)
		\\
		 \Phi_{{\rm s}yx}^{\alpha \beta}(\Im \omega_\lambda)&=-eT
		 \sum_{n\bk}\langle \psi_\alpha | v_{{\rm s}y}|\psi_\beta \rangle
		 \langle \psi_\beta | v_x | \psi_\alpha \rangle
		 \scr{G}_\alpha (\Im \tve_n) \scr{G}_\beta (\Im \tve_{n-}),
		 \label{eq73}
\end{align}
	where $\alpha, \beta = \pm$ denotes the conduction $(+)$ and or valence $(-)$ bands, $\psi_\pm$ are their eigen functions, and $\scr{G}_\pm (\Im \ve_n)=\left( \Im \ve_n \mp E\right)^{-1}$.
	This form is identical to eq. (\ref{eqpre65}).
	Then we find
	\begin{align}
		\sigma_{{\rm s}yx}^{\rm intra}&=\sigma_{{\rm s}yx}^{++} + \sigma_{{\rm s}yx}^{--} = 0,
		\\
		\sigma_{{\rm s}yx}^{\rm inter}&=\sigma_{{\rm s}yx}^{+-} + \sigma_{{\rm s}yx}^{-+} = 
		-\frac{em|\gamma|}{4\pi^2}
		(K_{{\rm s}yx}^{\rm I} + K_{{\rm s}yx}^{\rm II}).
\end{align}
	The SHC only comes from the interband contribution, $\sigma_{{\rm s}yx}^{\rm inter}$, and  intraband contribution, $\sigma_{{\rm s}yx}^{\rm intra}$, is completely zero.
	Consequently, at least for the Dirac electrons, the SHE is generated only by the interband effect, which is hardly affected by the impurity scattering.
	In previous sections, we have argued the interband effects of a {\it magnetic field} on the diamagnetism and the Hall conductivity.
	However, the present interband effect on the SHC is not due to the magnetic field.
	Only the interband matrix elements originates from the spin-orbit interaction give rise to the interband effects on the SHE.
	%

\subsection{Implications for experiments on Bi$_{1-x}$Sb$_x$}
\begin{figure}
\begin{center}
\includegraphics[width=7cm, bb=0 0 510 549]{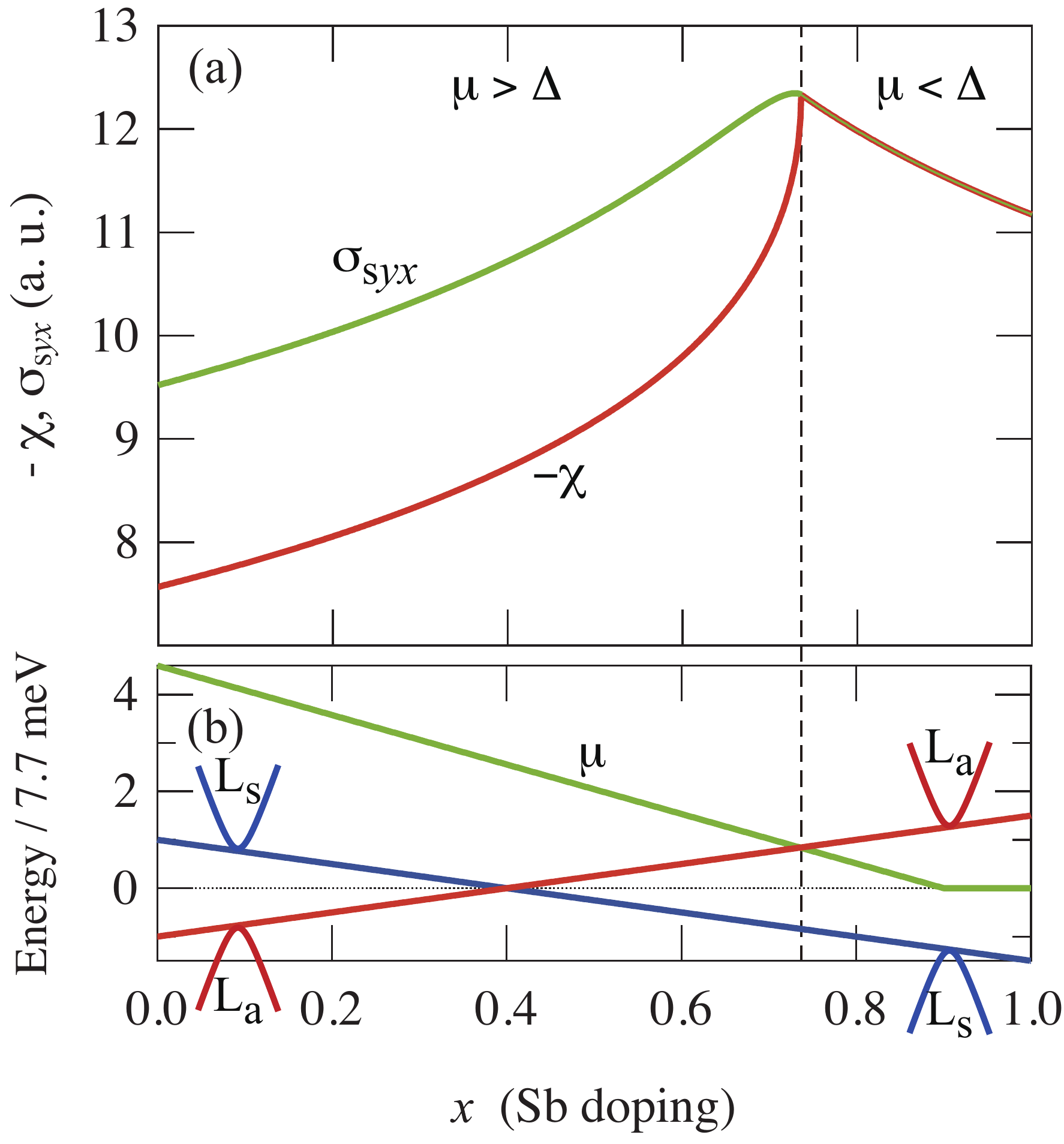}
\end{center}
\caption{(Color online) Antimony doping, $x$, dependence of (a) $\sigma_{{\rm s}yx} (x)$ and $-\chi(x)$, and (b) electron bands at $L$-point and $\mu$ of bismuth. The hole band at $T$-point is not shown. The approximation function of $\D (x)$ and $\mu (x)$ are given in the text.
}
\label{BiSb}
\end{figure}
	%
	Here we discuss implications of the present theoretical results based on the isotropic velocity for the experiments by taking into account the band structure of bismuth and and its alloy with antimony, in order to see the overall trends. 
	By substituting bismuth with antimony (Bi$_{1-x}$Sb$_x$), the band structure is changed as depicted in Fig. \ref{BiSb} (b).
	The valence band at $T$-point (not shown in Fig. \ref{BiSb}) is lowered by this substitution, resulting in the decrease of $\mu$.
	The doping dependences of $\D$ and $\mu$ can be simulated by\cite{Fuseya2012b}
	\begin{align}
		\pm \D (x) &= 1 - x/0.04,\\
		\mu(x) &=  4.6 - 4.6x/0.09.
\end{align}
	Substituting these doping dependence into eqs. (\ref{KIclean}) and (\ref{KIIclean}), the doping dependence of $\sigma_{{\rm s}yx} (x)$ and $\chi (x)$ are obtained as in Fig. \ref{BiSb} (a).

	The properties of $\chi(x)$ (the logarithmically increase and the kink structure at around $x\simeq 0.07$) agree quite well with the experimental results as is shown in Fig. \ref{Wehrli}.\cite{Verkin1967,Wehrli1968}
	The doping dependence of $\sigma_{{\rm s}yx}$ is expected to exhibit a similar behavior to $\chi(x)$, though $\sigma_{{\rm s}yx}$ will be slightly increased by the contribution from $K_{{\rm s}yx}^{\rm I}$.
	There is no anomaly at $x\simeq 0.04$, where the conduction and valence bands at $L$-points are inverted and the topology of the band is believed to change from trivial to non-trivial.\cite{Fu2007,Hsieh2008}
	Note that there should be finite contribution from the holes at $T$-point, but its magnitude would be much smaller than that from electrons at $L$-points, since the gap is much larger at $T$-point than that at $L$-points.

\section{SHE and Diamagnetism for the Wolff Hamiltonian}\label{Aniso}
	
	So far we have given the results of the Dirac Hamiltonian, eq. (\ref{isoWolff}), with isotropic effective velocity.
	But in actual materials, the electronic structure is anisotropic, which is not taken into account in the Dirac Hamiltonian.
	Especially, the electronic structure of bismuth is known to be highly anisotropic.
	Therefore, it is important to take into account the anisotropy of materials, when we were to compare the theoretical results to the experiments.
	Furthermore, it is crucially important to see whether the exact relationship between the SHC and orbital susceptibility, eq. (\ref{SHC_chi}), still holds even for the anisotropic case.
	Here we give results of the SHE and orbital magnetism for the Wolff Hamiltonian.\cite{Fuseya2014}

\subsection{SHE for the Wolff Hamiltonian}
	
	The spin magnetic moment of the Wolff Hamiltonian is defined by eq. (\ref{magnetic moment}).
	The velocity of the spin magnetic-moment (spin current) is then defined as 
	\begin{align}
	v_{{\rm s}j}^{i}&=\frac{\mu_{{\rm s}i} v_j + v_j \mu_{{\rm s}i}}{2\mu_{\rm B}}
	\nonumber\\
	&
	=\frac{1}{\mu_{\rm B}}\frac{\hbar e \Omega}{2c\D}
	\sum_{\lambda \mu \nu}\epsilon_{\lambda \mu \nu}
	\begin{pmatrix}
		0 & Q_i(\lambda)W_j(\mu) \sigma_\nu \\
		Q_i(\lambda) W_j (\mu) \sigma_\nu & 0
	\end{pmatrix},
\end{align}
	where $\bm{v}=\hbar^{-1}\partial \scr{H}/\partial \bk$ is the velocity operator.
	The operator $v_{{\rm s}j}^i$ is a tensor operator with respect to the direction of the spin magnetic-moment $i$ and that of the spin current $j$.

	The spin Hall conductivity can be obtained in the same manner as in the preceding section:
	\begin{align}
		\Phi_{{\rm s}jk}^i (\Im \omega_\lambda)&= -eT\sum_{n, \bk}
		{\rm Tr} \left[
	\scr{G}(\Im \tve_n) v_{{\rm s}j}^i \scr{G}(\Im \tve_{n-})v_k
	\right]
	\nonumber\\
	&
	=
	-eT
	\sum_{n, \bk}
	\frac{ 4\Im m\Omega (\Im \tilde{\ve}_n - \Im \tilde{\ve}_{n-})}{\{ (\Im \tilde{\ve}_n)^2 -E^2 \}\{ (\Im \tilde{\ve}_{n-})^2 -E^2 \}}
	\nonumber\\
	&\times
	\left[
	\sum_{\lambda \mu \nu} \epsilon_{\lambda \mu \nu} Q_i (\lambda) W_j (\mu) W_k (\nu)
	\right],
	\nonumber\\
	&=-e \epsilon_{kji}
	\frac{\D^2 \left(\alpha_{jk}^2 - \alpha_{jj}\alpha_{kk}\right)}{\sqrt{\D^3 \det \bm{\alpha}}}
	\nonumber\\
	&\times T \sum_{n, \tilde{k}_\mu}
	\frac{4\Im m  (\Im \tve_n - \Im \tve_{n-})}{\{ (\Im \tilde{\ve}_n)^2 -E^2 \}\{ (\Im \tilde{\ve}_{n-})^2 -E^2 \}},
	\label{eq137}
\end{align}
	Here we used the relation
	\begin{align}
		\sum_{\lambda \mu \nu} \epsilon_{\lambda \mu \nu} Q_i (\lambda) W_j (\mu) W_k (\nu)
		=\frac{\D^2}{\Omega }\epsilon_{kji}
		\left(
		\alpha_{jk}^2 - \alpha_{jj}\alpha_{kk}
		\right),
\end{align}
	where $\alpha_{jk}^2 - \alpha_{jj}\alpha_{kk}$ corresponds to the Gaussian curvature of the energy dispersion.
	The factor $\sum_{\lambda \mu \nu} \epsilon_{\lambda \mu \nu} Q_i (\lambda) W_j (\mu) W_k (\nu)$ gives finite contributions only when $i, j, k$ are perpendicular to each other.
	Figure \ref{Illust_SHE} shows the direction of the electric field, the spin current and the spin magnetic-moment.
	This configuration agrees with the phenomenological discussion given by Hirsch.\cite{Hirsch1999}
	The contributions of $\sum_{\lambda \mu \nu} Q_i (\lambda) W_j (\mu) W_k (\nu)$ for the other combinations, say, $i \, || \,k$ will vanish if we take into account total contributions from the whole Brillouin zone.\cite{Fuseya2014}

	When we compare eq. (\ref{eq137}) with eq. (\ref{eq65}), it is clear that the anisotropy is expressed only by the term $\epsilon_{kji}(\alpha_{jk}^2 - \alpha_{jj}\alpha_{kk})/\sqrt{\D^3 \det \hat{\alpha}}$.
	Finally, we obtain the spin Hall conductivity for the Wolff Hamiltonian:
	\begin{align}
		\sigma_{{\rm s}jk}^i =\frac{me}{4\pi^2 \hbar^2}
		\frac{\epsilon_{kji}\left(\alpha_{jk}^2 - \alpha_{jj}\alpha_{kk} \right)}{\sqrt{\det \hat{\alpha}/\D}}
		\left[ K_{{\rm s}yx}^{\rm I} + K_{{\rm s}yx}^{\rm II} \right],
\end{align}
	where $K_{{\rm s}yx}^{\rm I}$ and $K_{{\rm s}yx}^{\rm II}$ are the same functions as eqs. (\ref{KI}) and (\ref{KII}).
	Consequently, the properties of the SHE for the Wolff Hamiltonian are essentially the same as that for the Dirac Hamiltonian shown in Fig. \ref{SHE_mu}.

\subsection{Relationship between SHE and diamagnetism}	
	
	The orbital susceptibility for the Wolff model can be also calculated by eq. (\ref{Fukuyama formula}) in the same manner as that for the Dirac Hamiltonian.
	The calculations are more complicated than that for the SHE, but we obtain the following results after lengthy calculations:\cite{Fuseya2014}
	\begin{align}
		\chi^i = -\frac{e^2}{12\pi^2 \hbar c^2} 
		\frac{\alpha_{jk}^2 -\alpha_{jj}\alpha_{kk}}{\sqrt{\det \hat{\alpha}/\D}} K_{{\rm s}yx}^{\rm II},
\end{align}
	where the direction of the magnetic field is along $i$-direction, and $j$ and $k$ are perpendicular to the magnetic field.
	Here the orbital susceptibility is also given in terms of $K_{{\rm s}yx}^{\rm II}$, the same function as the isotropic case, and the anisotropy is expressed in terms of the Gaussian curvature.
	Now, we obtain an exact relationship between $\sigma_{{\rm s}jk}^i$ and $\chi^i$ in the insulating region of the Wolff Hamiltonian in the form:
	\begin{align}
		\sigma_{{\rm s}jk}^i = \frac{3m c^2}{\hbar e} \epsilon_{ijk} \chi^i.
		\label{SHE_chi}
\end{align}
	Surprisingly, the relationship for the isotropic case is also valid for the anisotropic case.
	The Wolff Hamiltonian is an effective Hamiltonian common to the Dirac electrons in solids, so that eq. (\ref{SHE_chi}) is a quite general relationship for Dirac electron systems whose Fermi level is located in the band gap.

\begin{figure}
\begin{center}
\includegraphics[width=4cm, bb=0 0 242 213]{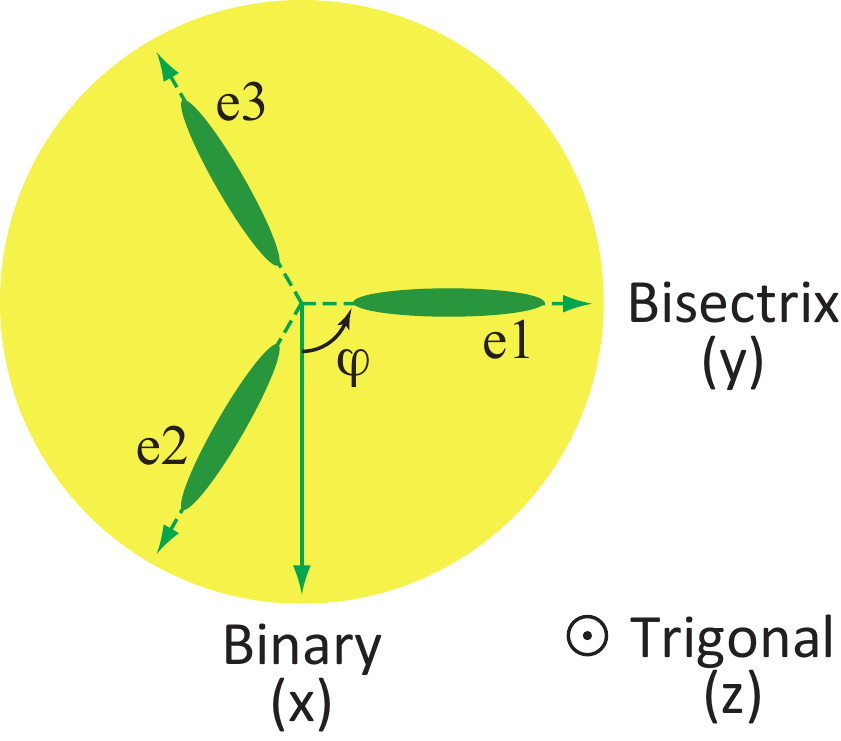}
\end{center}
\caption{(Color online) 
Illustration of three electron ellipsoids at $L$-points of bismuth.
}
\label{3FS}
\end{figure}
\subsection{Evaluation for bismuth}
	Here we give quantitative evaluations of the orbital susceptibility and the spin Hall conductivity for bismuth.
	The three electron ellipsoids at $L$-points of bismuth are displayed in Fig. \ref{3FS}.
	We label the ellipsoid along the bisectrix axis as ``e1", and the other (obtained by $\pm 120^\circ$ rotation about the trigonal axis) as ``e2" and ``e3".
	The inverse mass-tensor of e1 is given by\cite{Smith1964,ZZhu2011,Dresselhaus1971,Issi1979}
	\begin{align}
		\hat{\alpha}_{\rm e1} = 
		\begin{pmatrix}
			\alpha_1 & 0 & 0 \\
			0 & \alpha_2 & \alpha_4 \\
			0 & \alpha_4 & \alpha_3
		\end{pmatrix}.
\end{align}
	For e2 and e3, we have
	\begin{align}
		\hat{\alpha}_{\rm e2, e3} = \frac{1}{4}
		\begin{pmatrix}
			\alpha_1 + 3 \alpha_2 & \pm \sqrt{3}(\alpha_1 - \alpha_2) & \pm 2\sqrt{3} \alpha_4 \\
			 \pm \sqrt{3}(\alpha_1 - \alpha_2)  & 3\alpha_1 + \alpha_2 & -2 \alpha_4 \\
			 \pm 2\sqrt{3} \alpha_4 & -2 \alpha_4 & 4 \alpha_3
		\end{pmatrix}.
\end{align}
	The total contribution from three electron ellipsoids is given by the summation of the each Gaussian curvature:
	\begin{align}
		\left(\alpha_{xy}^2 - \alpha_{xx}\alpha_{yy}\right)_{\rm total} &= -3 \alpha_1 \alpha_2 \equiv \kappa_{\parallel},\\
		\left(\alpha_{yz}^2 - \alpha_{yy}\alpha_{zz}\right)_{\rm total} &= \frac{3}{2}\alpha_4^2 -\frac{3}{2}(\alpha_1 + \alpha_2) \alpha_3 \equiv \kappa_{\perp},\\
		\left(\alpha_{zx}^2 - \alpha_{zz}\alpha_{xx}\right)_{\rm total} &= \frac{3}{2}\alpha_4^2 -\frac{3}{2}(\alpha_1 + \alpha_2) \alpha_3 =\kappa_{\perp},
\end{align}
	where $x$-, $y$- and $z$-axis are taken to be binary-, bisectrix- and trigonal-axis.
	The Gaussian curvature for the binary plane is the same as that for the bisectrix plane.
	%
\begin{table}
\caption{Parameters for the mass tensor $\hat{m}$\cite{ZZhu2011} (in the unit of the bare electron mass $m$) and the inverse mass tensor $\hat{\alpha}$ (in the unit of $m^{-1}$).}
\label{t4}
\begin{center}
\begin{tabular}{lcccc}
\hline
\hline
$i$ & $1$ & $2$ & $3$ & $4$ \\
\hline
mass ($m_i$) & 0.00124 & 0.257 & 0.00585 & -0.0277 \\
inverse mass ($\alpha_i$) & 806 & 7.95 & 349 & 37.6 \\
\hline
\end{tabular}
\end{center}
\end{table}

	The values of $\alpha_1 \sim \alpha_4$ have been experimentally determined\cite{Smith1964,ZZhu2011}, which are listed in Table \ref{t4}.
	Using these values, we can evaluate the total Gaussian curvature as follows:
	\begin{align}
		\kappa_{\parallel} m^2&= -1.92 \times 10^4 ,\\
		\kappa_{\perp} m^2 &= -4.24 \times 10^5.
\end{align}
	The total Gaussian curvature is quite anisotropic.
	For $H\perp z$, the contribution of electron is much larger than that of holes, so that $\chi_\perp \simeq \chi_\perp^{\rm (e)}$.\cite{Fukuyama1970}
	For $H\parallel z$, on the other hand, the contribution of electrons is smaller than that of holes, so that we need to evaluate the hole contribution for this direction.
	Unfortunately, however, it is rather difficult to give an accurate estimation of hole contribution, since the magnitude of the energy gap at $T$-point is still unclear.
	(It is estimated to be at least larger than 200 meV.\cite{Golin1968,Bate1969,Verdun1976})
	Hence, we shall give the estimation only for $H\perp z$ below.
	The contribution of electrons to the orbital susceptibility at zero temperature is evaluated as
	\begin{align}
		\\
		\chi_{\perp}^{\rm (e)}&=
		-\frac{e^2 \kappa_\perp}{12\pi^2 \hbar c^2} 
		\sqrt{\frac{\D_{\rm e}}{\det \hat{\alpha}}}
		K_{{\rm s}yx}^{\rm II}(\mu_{\rm e})
		\nonumber\\
		&=
		\left( 9.27-6.11\times \ln \frac{E_{\rm c}}{\D_{\rm e}} \right) \times 10^{-6} \, \mbox{emu},
\end{align}
	with the band gap $\D_{\rm e}=7.65$ meV and the chemical potential $\mu_{\rm e}=35.3$ meV for pure Bi.\cite{Smith1964,ZZhu2011}
	The magnitude of $\chi_\perp^{\rm (e)}$ is of the order of $10^{-5}$emu.
	If we choose $E_{\rm c}/\D=100$, we have $\chi_\perp =-1.89 \times 10^{-5}$emu.
	This agrees well with the experimental results $\chi_\perp =-1.94 \times 10^{-5}$emu obtained by Otake {\it et al.}\cite{Otake1980}.

	The spin Hall conductivity at zero temperature is evaluated as
	\begin{align}
		e\sigma_{{\rm s}\perp}^{\rm (e)}&=\frac{me^2 \kappa_\perp}{4\pi^2 \hbar^2}\sqrt{\frac{\D_{\rm e}}{\det \hat{\alpha}}}
		\left[K^{\rm I}(\mu_{\rm e}) + K^{\rm II}(\mu_{\rm e}) \right]
		\nonumber\\
		&=\left(-0.855+1.58\times \ln \frac{E_{\rm c}}{\D_{\rm e}}\right)\times 10^4 \, \Omega^{-1}{\rm cm}^{-1}
\end{align}
	The magnitude of $e\sigma_{{\rm s}\perp}^{\rm (e)}$ is of the order of $10^4 \,\Omega^{-1}{\rm cm}^{-1}$.
	For $E_{\rm c}/\D =100$, we have $e|\sigma_{{\rm s}\perp}^{\rm (e)}|= 6.44 \times 10^4 \,\Omega^{-1}{\rm cm}^{-1}$.

	Also, we can estimate $e\sigma_{{\rm s}xy}$ directly from the experimental value of the orbital susceptibility by using the formula eq. (\ref{SHE_chi}):
	\begin{align}
		e\sigma_{{\rm s}xy}= (2.59 \times 10^9 ) \chi \,\Omega^{-1}{\rm cm}^{-1},
		\label{eq135}
\end{align}
	since $3mc^2/\hbar =2.59 \times 10^9\, \Omega^{-1}{\rm cm}^{-1}$.
	The contribution of $K_{{\rm s}yx}^{\rm I}$ is neglected in this formula, so that it is valid only in the insulating region.
	However, the general situation will not vary from the results obtained by eq. (\ref{eq135}), since $K_{{\rm s}yx}^{\rm I}$ gives almost uniform contribution with respect to $\mu$, and it is smaller than $K_{{\rm s}yx}^{\rm II}$.

	The magnetic susceptibility of pure bismuth at room temperature is $\chi_{\perp}^{\rm (rt)}=1.43 \times 10^{-5}$emu, so that we obtain $e|\sigma_{{\rm s}\perp}^{\rm (rt)}|=3.70 \times 10^4 \,\Omega^{-1}{\rm cm}^{-1}$ at room temperature.
	The spin Hall conductivity of Pt is estimated as $e|\sigma_{{\rm s}\perp}^{\rm (rt)}|\simeq 2.4 \times 10^2 \,\Omega^{-1}{\rm cm}^{-1}$, which is about $10^4$ times larger than the value reported in $n$-type semiconductors.\cite{Kimura2007,Guo2008,Kontani2008}
	Therefore, the obtained value for bismuth of the order of $ 10^4 \,\Omega^{-1}{\rm cm}^{-1}$ is quite large compared to the SHC of conventional systems.
	The spin Hall conductivity of bismuth would be further enhanced by alloying with Sb, since the increase of the diamagnetism has been observed.\cite{Wehrli1968}

\section{Spin-Polarized Electric Current}\label{Sec_SPEC}

	It is important that the wave function of the Dirac electron systems under a magnetic field is explicitly given and then the transport phenomena of Dirac electrons under a magnetic field can be studied in a rigorous manner.
	The pioneering work of the magneto-optical properties of Dirac electrons was established by Wolff in 1964.\cite{Wolff1964}
	Wolff pointed out that the unique feature of bismuth is that the spin transition by an external electric filed is possible due to the spin-orbit interaction in addition to the ordinary orbital transition.
	The problem is that we cannot control the spin transition only, since the orbital and spin quantum number are completely mixed as is discussed in \S\ref{Sec_magnetic}.
	However, it it possible to generate a spin-polarized electric current by using the circularly polarized light.\cite{Fuseya2012a}
	The key of this possibility is due to the lowest energy level, where the spin state is uniquely specified.
	%

\subsection{Magneto-optical conductivities}	
\begin{figure}
\begin{center}
\includegraphics[width=4cm, bb=0 0 465 1259]{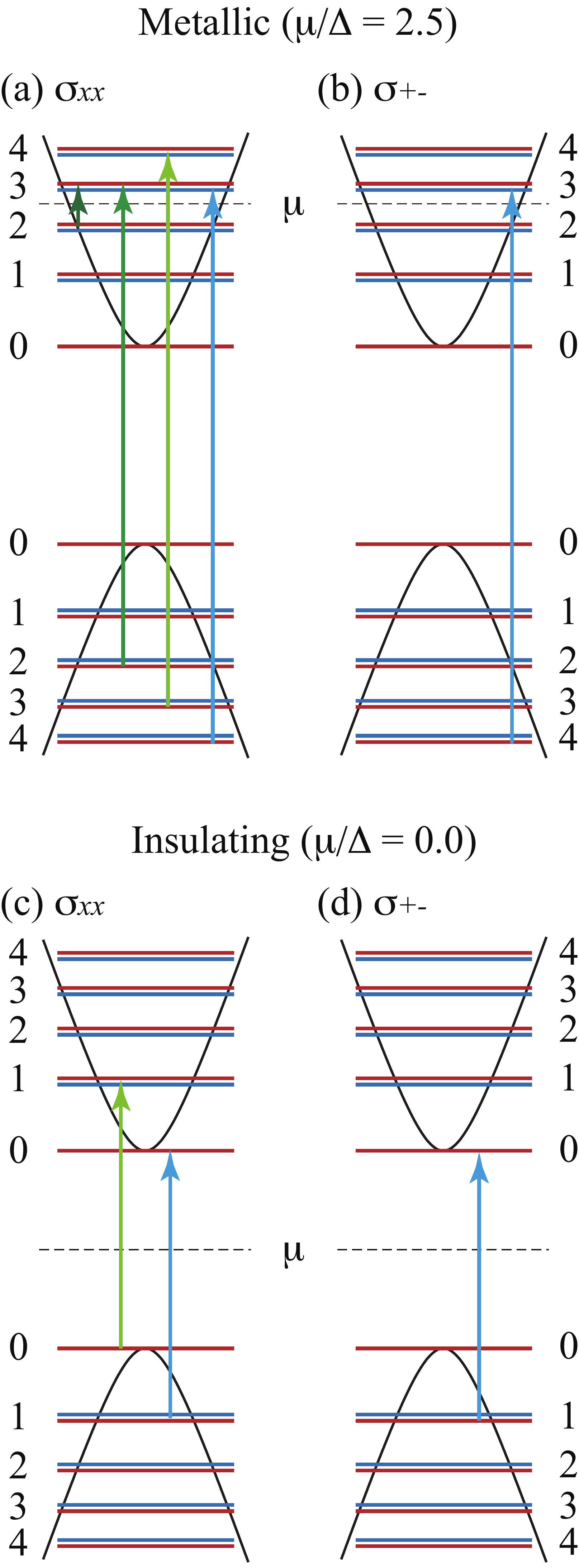}
\end{center}
\caption{(Color online) 
Energy levels under a magnetic field and possible transitions for $\sigma_{xx}$ and $\sigma_{+-}$ in the metallic region ($\mu/\D=2.5$; (a), (b)) and for the insulating region ($\mu/\D=0$; (c), (d)), respectively.
The numbers denotes the total angular momentum $j$.
}
\label{illust_SPEC}
\end{figure}
	%
	
	First we briefly explain the general properties of the magneto-optical conductivities.
	The diagonal conductivity and the Hall conductivity for the Dirac Hamiltonian, eq. (\ref{isoWolff}), under finite magnetic field are obtained by the same way as \S\ref{Sec_Hall} and \S\ref{Sec_SHE} in the forms:
\begin{align}
		&\sigma_{xx}(\omega)=\frac{e^2 N_{\rm L}(H_z)}{\Im \omega}
		\sum_{j=0}^\infty \sum_{k_z, \pm}
		\frac{\gamma^2}{2E_j E_{j+1}}
		\nonumber\\
		&\times
		\Biggl[
		(E_j E_{j+1} - \D^2 - \gamma^2 k_z^2 )
		F_{+}(\pm E_j , \pm E_{j+1})
		\nonumber\\
		&+(E_j E_{j+1} + \D^2 + \gamma^2 k_z^2 )
		F_{+}(\pm E_j , \mp E_{j+1})
		\Biggr],
		\label{Sxx}
		\\
		&\sigma_{yx}(\omega)=\frac{e^2 N_{\rm L}(H_z)}{\Im \omega}
		\sum_{j=0}^\infty \sum_{k_z, \pm}
		\frac{\Im \gamma^2}{2E_j E_{j+1}}
		\nonumber\\
		&\times
		\Biggl[
		(E_j E_{j+1} - \D^2 - \gamma^2 k_z^2 )
		F_- (\pm E_j , \pm E_{j+1}) 
		\nonumber\\
		&+(E_j E_{j+1} + \D^2 + \gamma^2 k_z^2 )
		F_- (\pm E_j , \mp E_{j+1}) 
		\Biggr],
		\label{Sxy}
\end{align}
	where $E_j$ is given by eq. (\ref{ene}) with $j=n +1/2 \pm 1/2$, and $N_{\rm L}(H)=eH/2\pi c$ is the degeneracy of a Landau level per the unit area perpendicular to the field.
	The contribution from the Green's function is expressed by the function
	\begin{align}
		&F(E_\alpha, E_\beta)=\frac{1}{2\pi \Im}
		\int_{-\infty}^{\infty}\!\! d\ve f(\ve)
		\Biggl[
		\frac{1}{\ve+\omega -E_\beta + \Im \Gamma}\frac{1}{\ve -E_\alpha + \Im \Gamma}
		\nonumber\\
		&
		-\frac{1}{\ve+\omega -E_\beta + \Im \Gamma}\frac{1}{\ve -E_\alpha - \Im \Gamma}
		+\frac{1}{\ve -E_\beta + \Im \Gamma}\frac{1}{\ve -\omega -E_\alpha - \Im \Gamma}
		\nonumber\\
		&
		-\frac{1}{\ve -E_\beta - \Im \Gamma}\frac{1}{\ve -\omega -E_\alpha - \Im \Gamma}
		\Biggr],
		\label{F function}
\end{align}
	 and $F_{\pm}(E_\alpha, E_\beta) = F(E_\alpha, E_\beta) \pm F(E_\beta, E_\alpha)$.
	 The chemical potential $\mu$ is included in $f(\ve)$.
	%
\begin{figure}
\begin{center}
\includegraphics[width=6.0cm,bb=0 0 537 1637]{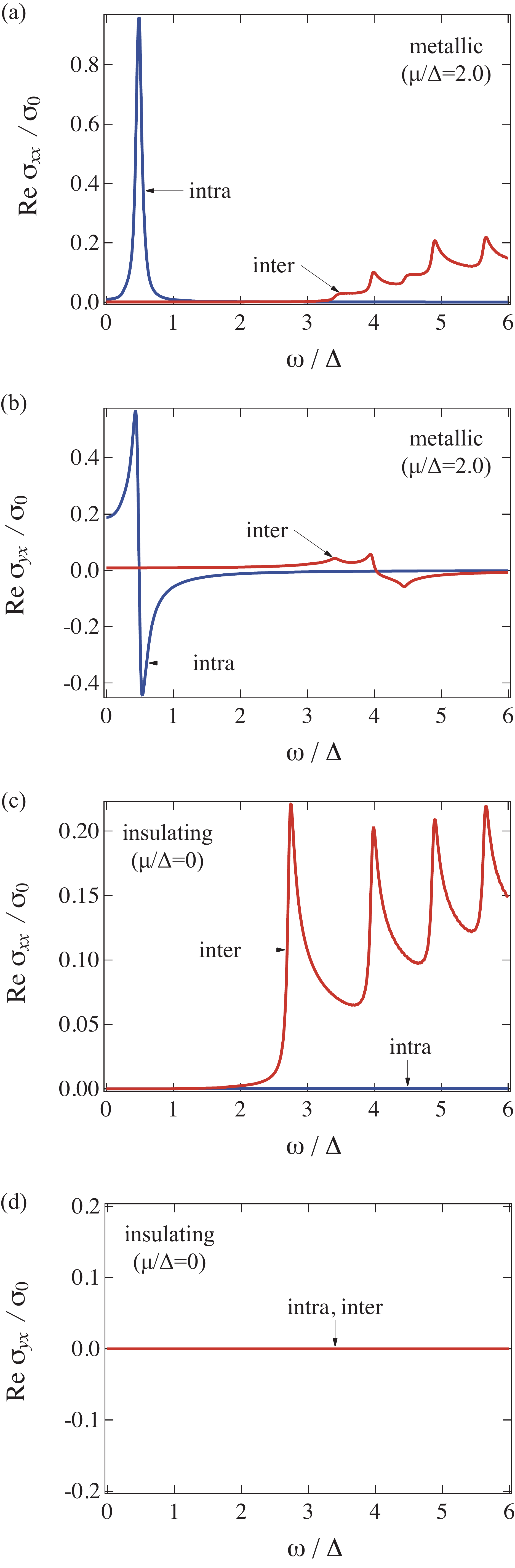}
\end{center}
\caption{(Color online) Frequency dependence of (a) Re $\sigma_{xx} (\mu/\D=2.0)$, (b) Re $\sigma_{yx}(\mu/\D=2.0)$,
(c) Re $\sigma_{xx} (\mu/\D=0)$ and (d) Re $\sigma_{yx}(\mu/\D=0)$ for $\wc /\D= 1.0$.
}
\label{CxxCxy}
\end{figure}
	%

	From these results we see the selection rule of both $\sigma_{xx}$ and $\sigma_{xy}$:
	The transition with $\D j= \pm 1$ is allowed and the final state ($E_\beta$) must locate at the opposite side of $\mu$ from the initial state ($E_\alpha$), wherever $\mu$ is, e.g., $E_\alpha <\mu<E_\beta$.
	(Here we only consider the case with $\omega >0$.)
	The possible transitions for $\sigma_{xx}(\omega)$ is illustrated in the left panels of Fig. \ref{illust_SPEC} (a) and (c).
	The frequency dependences of $\sigma_{xx}(\omega)$ and $\sigma_{xy}(\omega)$ are shown in Fig. \ref{CxxCxy}, where the magnetic field is set to $\wc/\D=1.0$, i.e., the energy level splitting is comparable to the band gap.
	In order to obtain such a wide energy level splitting, we need $B\gtrsim 600$T for ordinary semiconductors (e.g., $\D=10^3$meV and $\mc /m$=0.07).
	On the other hand, in the case of bismuth ($\D=7$meV and $\mc/m=0.01$), we only need $B=0.60$T, which can be easily achieved.

\begin{figure}
\begin{center}
\includegraphics[width=70mm, bb=0 0 566 868]{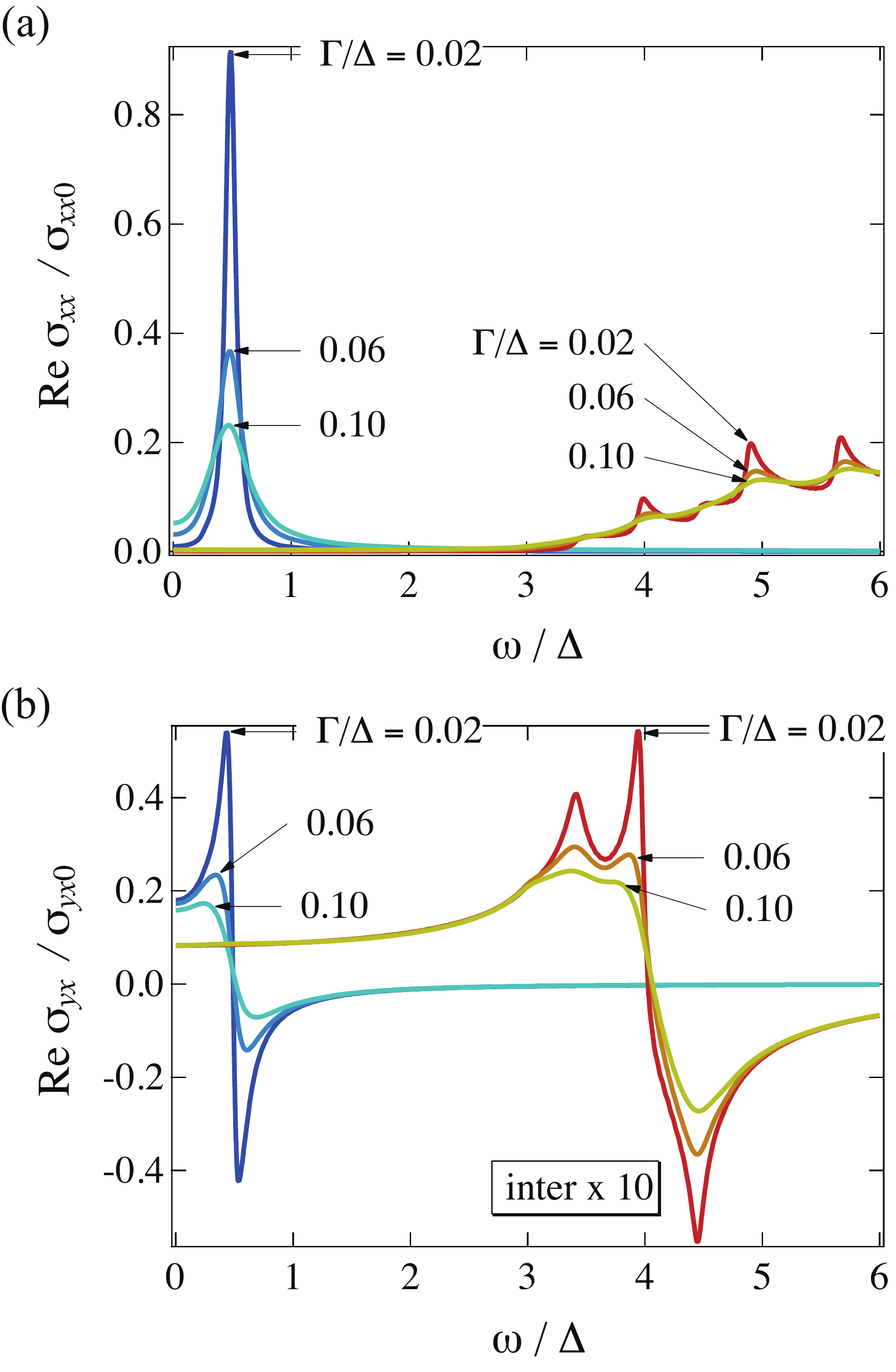}
\end{center}
\caption{(Color online) 
Impurity scattering dependence of (a) $\sigma_{xx}(\omega)$ and (b) $\sigma_{xy}(\omega)$ for $\wc /\D =1.0$ and $\mu/\D=2.0$.
$\sigma_{xx}^{\rm inter}$ is amplified in order to make the comparison clear.
}
\label{Cxx_impurity}
\end{figure}

	For the metallic case, $|\mu| \gg \D$, the sharp peak in $\omega <\D$ originates from the intraband transition, while the saw-toothed structure in $\omega > 2\D$ originates from the interband transitions.
	The interband contributions are clearly separated in the frequency space, whereas they are completely mixed for dc conductivities.
	Therefore, by tuning the frequency, we can selectively take out $\sigma_{xy}^{\rm inter}$, which is discussed in \S\ref{Sec_Hall}.
	Moreover, the magnitude of the interband contribution $\sigma_{\mu\nu}^{\rm inter}$ can be comparable to the intraband one $\sigma_{\mu\nu}^{\rm intra}$ for finite frequency, though $\sigma_{\mu\nu}^{\rm inter} \ll \sigma_{\mu\nu}^{\rm intra}$ for dc conductivities as is shown in Fig. \ref{results}.
	The impurity scattering $\Gamma$ dependences of $\sigma_{xx}$ and $\sigma_{xy}$ are shown in Fig. \ref{Cxx_impurity}.
	The intraband part is strongly reduced by $\Gamma$, whereas the interband part is not affected so much, which is consistent with the discussion in \S\ref{Sec_Hall}.
	%

\subsection{100\% Spin-Polarized Electric Current (SPEC)}

	%
	Now we explain a new mechanism to generate a fully spin-polarized electric current (SPEC) by the use of the circularly polarized light.\cite{Fuseya2012a}
	This mechanism utilizes the specific property of the lowest energy level (the $j=0$ level).

	For $\sigma_{xx}(\omega)$ in the metallic state, $|\mu|\gg\D$, it is prohibited by the selection rules to excite electrons into the $j=0$ level (Fig. \ref{illust_SPEC} (a)).
	In the insulating state, there are two processes for the smallest excitation energy involving $j=0$ state (Fig. \ref{illust_SPEC} (c)).
	In this case, we do not have net spin polarization, since the process includes $j=1$ state, where spin up and down states are degenerate.
	However, the situation completely changes if we see the response to the circularly polarized light.

	The response to the circularly polarized light is calculated in terms of $v_\pm \equiv (v_x \pm \Im v_y)/\sqrt{2}$.
	Then the conductivity for the circularly polarized light is given by
	\begin{align}
		\sigma_{+-}(\omega) &= \sigma_{xx}(\omega) + \Im \sigma_{yx}(\omega)
		\nonumber\\
		&
		=
		\frac{e^2 N_{\rm L}(H_z)}{\Im \omega}
		\sum_{j=0}^\infty \sum_{k_z, \pm}
		\frac{\gamma^2}{2E_j E_{j+1}}
		\nonumber\\
		&\times
		\Biggl[
		(E_j E_{j+1} - \D^2 - \gamma^2 k_z^2 )
		 F(\pm E_{j+1}, \pm E_j)
		\nonumber\\
		&+(E_j E_{j+1} + \D^2 + \gamma^2 k_z^2 )
		F(\mp E_{j+1}, \pm E_j) 
		\Biggr].
		\label{Spm}
\end{align}
	In this case, the possible transition is restricted to $\D j=-1$.
	(On the contrary, the transition only with $\D j=+1$ is possible for $\sigma_{-+}(\omega)$.)
	Therefore, in the insulating state, only the transition shown in Fig. \ref{illust_SPEC} (d) is allowed for the smallest excitation energy, which leads to the net spin polarization.
	%
\begin{figure}
\begin{center}
\includegraphics[width=6cm, bb=0 0 544 895]{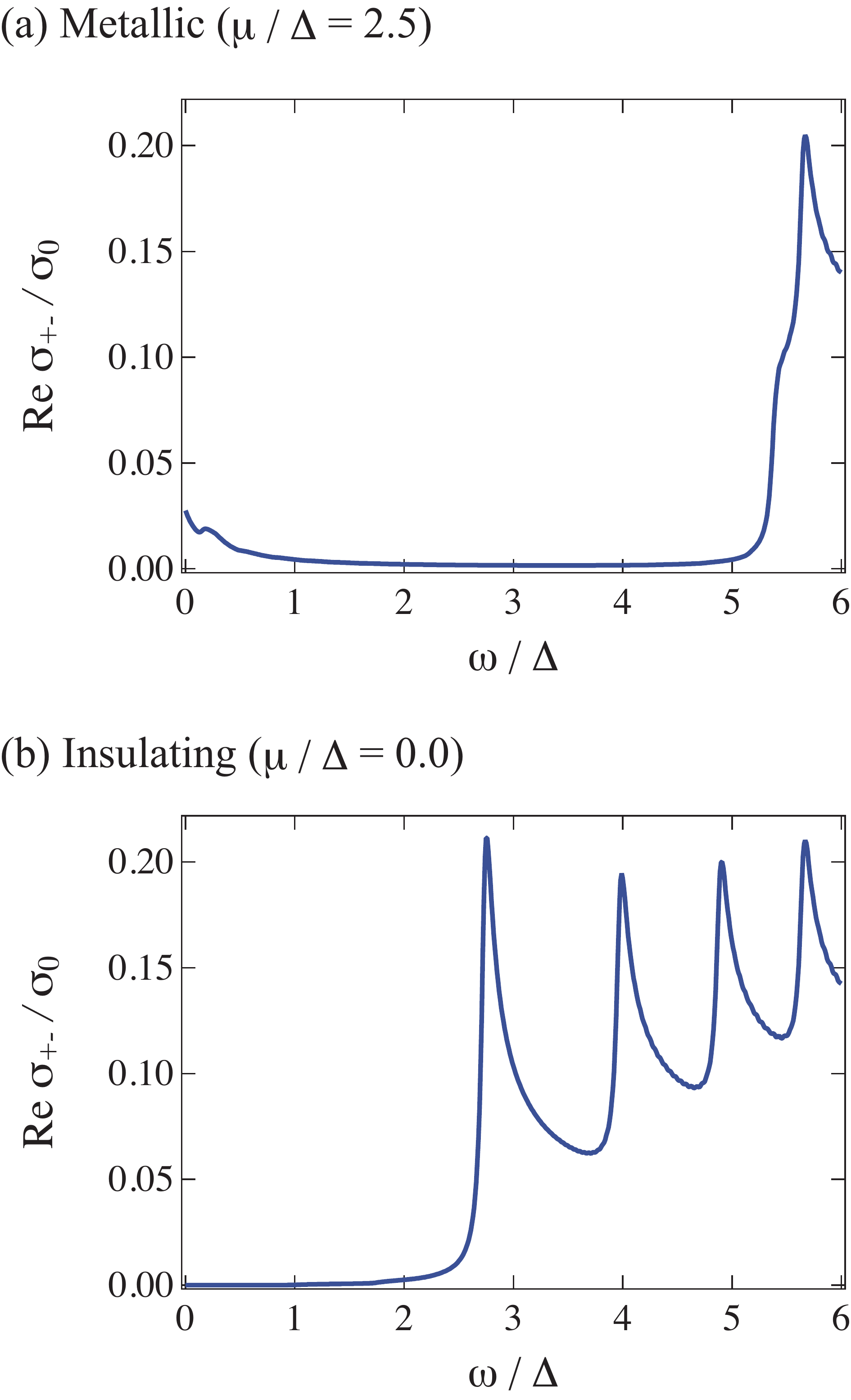}
\end{center}
\caption{(Color online) 
Frequency dependences of $\sigma_{xx}$ (left) and $\sigma_{+-}$ (right) for (a) insulating ($\mu/\D = 0$), and (b) metallic ($\mu/\D = 2.5$) regions with $\wc/\D = 1.0$.
}
\label{SPEC_result}
\end{figure}
\begin{figure}
\begin{center}
\includegraphics[width=6cm, bb=0 0 550 826]{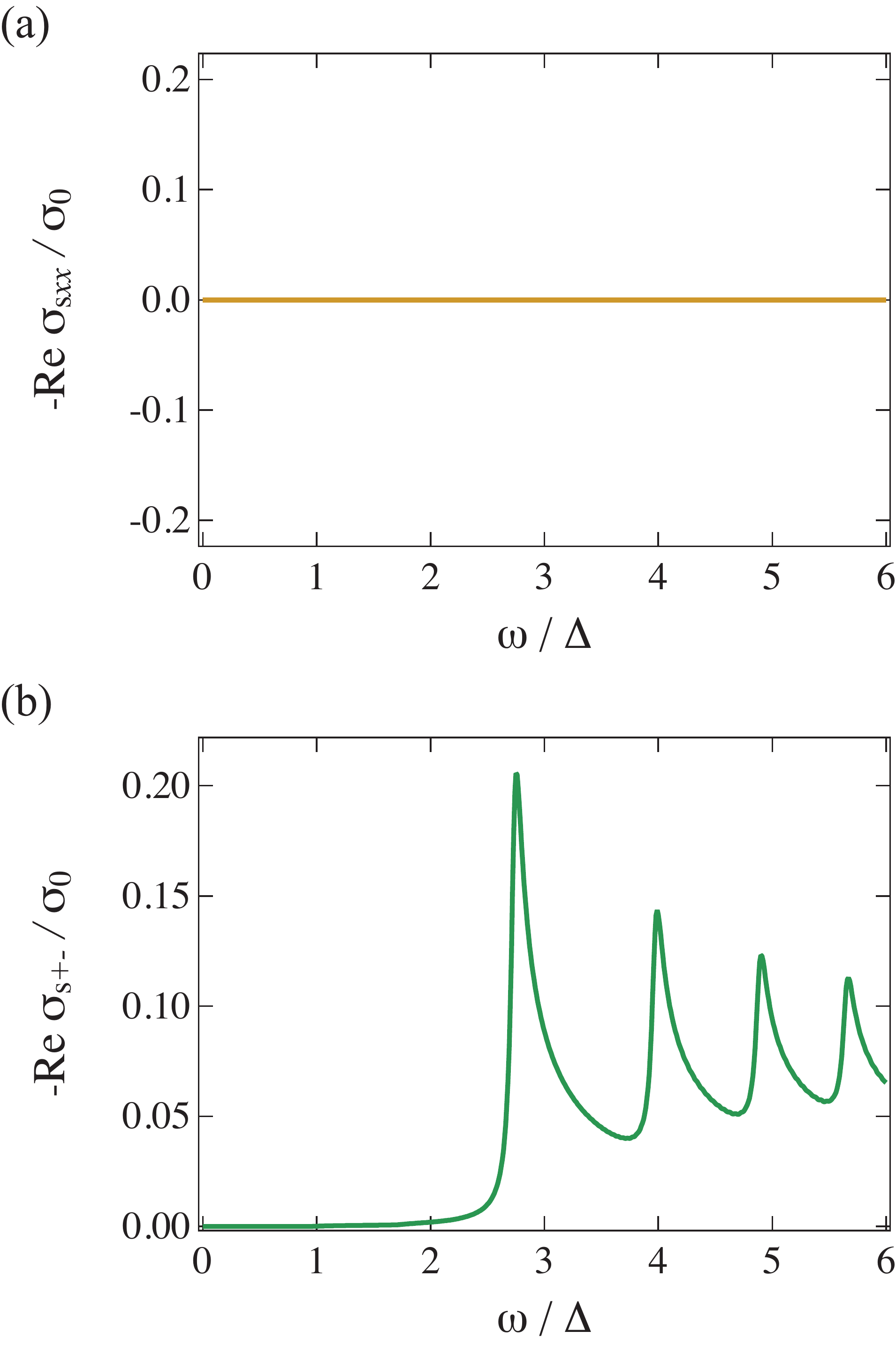}
\end{center}
\caption{(Color online) 
Frequency dependences of (a) $\sigma_{{\rm s}xx}$ and (b) $\sigma_{{\rm s}+-}$ (right) for insulating ($\mu/\D = 0$) regions with $\wc/\D = 1.0$.
}
\label{SPEC_spin}
\end{figure}
	%
	Its frequency dependences are shown in Fig. \ref{SPEC_result} for (a) metallic ($\mu/\D=2.5$) and (b) insulating ($\mu/\D = 0$) region.
	At first sight, there is no essential difference between $\sigma_{xx}(\omega)$ and $\sigma_{+-}(\omega)$, though the intraband transition is impossible for $\sigma_{+-}(\omega)$ when $\mu>\D$.
	However, their spin structures are completely different.
	This is confirmed by calculating the response of the spin-velocity operator (spin current) to the circularly polarized light, which is shown in Fig. \ref{SPEC_spin} for $\mu/\D=0$ and $\wc/\D = 1.0$.
	Here, the definition of the spin current is the same as that is discussed in \S\ref{Sec_SHE}.
	We see that the spin current is actually realized for $\sigma_{+-}$, whereas there is no spin current for $\sigma_{xx}$.
	The spin polarization is evaluated by $|\sigma_{{\rm s}\mu \nu}(\omega)|/\sigma_{\mu \nu}$.
	Figures \ref{SPEC_result} and \ref{SPEC_spin} indicate that $\sigma_{+-}(\omega)$ in the insulating region is highly spin-polarized, while $\sigma_{xx}(\omega)$ is not.
	Especially, at the lowest excitation energy $\omega_1$, $|\sigma_{{\rm s}+-} (\omega_1)|/\sigma_{+-}(\omega_1)=1$, i.e., we have 100\% spin polarization.	
	One may think that the spin relaxation is very fast in cases with a strong spin-orbit interactions, thereby destroying the spin-polarization.
	However, this is not the case for the present mechanism.
	The energy levels under consideration are the exact eigenstates of the Wolff Hamiltonian, which already includes the spin-orbit interaction, so that different energy levels are not mixed with each other. 
	Furthermore, the present SPEC uses the $j=0$ level, whose spin is unique and is not mixed, so that the spin-polarization is kept.

	This SPEC flows through the bulk, not the surface as in the topological insulators, and then is much easier to detect.
	In fact, the magneto-optical measurements on bismuth already exhibit clear peak structures.\cite{Maltz1970,Vecchi1974,Vecchi1976,Dresselhaus1971}
	The present proposal of SPEC will be confirmed experimentally, e.g., by the ordinary magneto-optical measurements with the use of the Kerr effect.\cite{Kato2004}

	When we compare the theoretical results for bismuth with the experiments, we need to consider the contribution from the holes at $T$-point in general.
	For the dc responses, $\omega \to 0$, the hole contribution is always mixed with the electron contributions, so it is necessary to take into account the hole contribution.
	For the ac responses, on the other hand, the peak structures due to holes will appear in much higher $\omega$ than that due to electrons, since the band gap at $T$-point ($\gtrsim 200$ meV\cite{Golin1968,Bate1969,Verdun1976}) is much larger than that at $L$-point ($\simeq 15$ meV).
	Thus we can discard the hole contribution for the present subject.
	%

\section{Summary}\label{Sec_Summary}
	In this paper, a review is given of recent progress on the transport phenomena, such as the weak-field Hall effect, spin Hall effect and ac conductivity, of Dirac electrons in bismuth. 
	The interband effect is the key behind these recent topics.

	The long-standing mystery of the large diamagnetism of bismuth had been solved by a careful analysis of the interband effect of a magnetic field in the presence of large spin-orbit interaction based on the Wolff Hamiltonian; the effective Hamiltonian of bismuth, which is essentially equivalent to the Dirac Hamiltonian, except for the anisotropic effective velocity.
	The Wolff Hamiltonian is a general effective Hamiltonian not only for bismuth, but also for various materials with a small band gap (i.e., a small effective mass) and a strong spin-orbit interaction.
	The interband effect of a magnetic field can generates large diamagnetic current, which is dissipationless, even in insulators.
	Although there has been some suggestions that this interband effect may also affect the transport phenomena, it has not been examined in detail for a long time.
	Recently, these interband effects and possibilities of the dissipationless transport phenomena have attracted much attention in various context, such as the Berry phase, spin current, and topological insulators.
	In this review, we have investigated the following transport phenomena, (i) weak-field Hall effect, (ii) spin Hall effect, and (iii) ac conductivity, based on the Wolff Hamiltonian, paying a special attention to the interband effect and the relationship between the transport phenomena and the diamagnetism.

	(i) The interband effect of a magnetic field on the weak-field Hall effect is studied based on the Dirac Hamiltonian in solids, which is isotropic version of the Wolff Hamiltonian.
	The interband contribution to the Hall conductivity $\sigma_{xy}^{\rm inter}$ takes its maximum value when the chemical potential $\mu$ locates at around the band-edge, and decreases as carrier increases.
	Also, $\sigma_{xy}^{\rm inter}$ is hardly affected by the impurity scatterings.
	These properties are quite different from those of ordinary transport coefficients.
	Instead, they share common features with the properties of orbital susceptibility $\chi$, which fact strongly suggests that the interband contribution to the Hall conductivity has a common origin with the diamagnetic current.

	(ii) The spin Hall effect is also investigated based on the Wolff Hamiltonian.
	It is found that the spin Hall conductivity is related to the orbital susceptibility by a simple and clear formula only with physical constants, $\sigma_{{\rm s}jk}^i=(3mc^2 /\hbar e)\epsilon_{ijk} \chi^i$, when $\mu$ locates in the band gap.
	($\sigma_{{\rm s}jk}^i$ is the conductivity tensor for the response of the velocity operator ($j$-direction) of the spin magnetic-moment ($i$-direction) to the electric field ($k$-direction), and $\chi^i$ is the orbital susceptibility under a magnetic field in $i$-direction.)
	There, the spin Hall current flows even in the insulating states without electric current, i.e., the dissipationless spin Hall insulator is achieved.
	We see the spin Hall effect is generated only by the interband matrix element of spin currents.

	Based on this theoretical finding, the magnitude of spin Hall conductivity is estimated for bismuth and its alloys with antimony in terms of experimental value of diamagnetism.
	The magnitude of spin Hall conductivity of bismuth turns out to be as large as $e\sigma_{{\rm s}xy}\sim 10^4\, \Omega^{-1}{\rm cm}^{-1}$, which is about 100 times larger than that of Pt. 
	Its magnitude will be further increased by alloying with antimony toward insulating state.

	(iii) A possible mechanism of spin-polarized electric current is proposed under a magnetic field.
	By using the circularly polarized light and tuning its frequency in insulating states, we can excite valence band electrons into the lowest energy level of conduction band, where spins of a particular direction can occupy.
	Hence 100\% spin-polarized magneto-optical current is achieved.
	%

\vspace{1cm}
\begin{acknowledgment}

\acknowledgment


The authors would like to thank the following people for fruitful discussions and comments: Y. Ando, K. Behnia, A. Collaudin, B. Fauqu\'e, T. Goto, H. Harima, J. Inoue, A. Kobayashi, H. Kohno, K. Miyake, E. Shikoh, M. Shiraishi, Y. Suzumura and Z. Zhu.
This  work is supported by Grants-in-Aid for Scientific Research on ``Dirac electrons in Solids" (No. 24244053), and by JSPS Bilateral Programs (No. 13428660).
Y. F. is also supported by Young Scientests B (No. 25870231).
	
\end{acknowledgment}

\appendix
\section{Definition of spin current} \label{Sec_Definition}

	When we calculate the spin Hall conductivity based on the linear response theory (Kubo formula)\cite{Kubo1957}, we need to define the spin current operator.
	Basically, the spin current should be defined by the product of the spin operator and the velocity operator as $j_\nu^\mu \equiv \left\{ v_\nu , s_\mu \right\}/2$.
	For example,  for the intrinsic SHE proposed for the $n$-type semiconductors with Rashba spin-orbit coupling, Sinova {\it et al.} assumed that the spin current operator is given as $j_i^z=\hbar\{ \sigma_z, v_i \}/4$.\cite{Sinova2004,note7} 
	For the intrinsic SHE proposed Luttinger Hamiltonian\cite{Murakami2003}, which is an effective Hamiltonian for the the $p$-type semiconductors of Si, Ge\cite{Luttinger1956} or GaAs\cite{note6}, Murakami {\it et al.} assumed the spin current operator is given as $j_i^\ell =(1/6)\{ v_i, S_\ell \}$, where $S_\ell$ is the spin-3/2 matrix.\cite{Murakami2004a}
	However, there are some problems with the spin current.
	First, the spin current is not conserved.
	Second, there does not exist a direct measurement of the spin current at the moment.
	In case of the charge-current, there is a conservation law between the charge density $\rho_{\rm c}$ and the charge current $\bm{j}_{\rm c}$:
	\begin{align}
		\frac{\partial \rho_{\rm c}}{\partial t} + \bm{\nabla}\cdot \bm{j}_{\rm c}=0,
\end{align}
	i.e., the continuity equation.
	In case of the spin current, on the other hand, the spin density $\rho_{\rm s}$ and the spin current $\bm{j}_{\rm s}$ satisfy the non-conserved equation:
	\begin{align}
		\frac{\partial \rho_{\rm s}^\alpha}{\partial t} + \bm{\nabla}\cdot \bm{j}_{\rm s}^\alpha =\mathcal{T}^\alpha,
		\label{nonconserve}
\end{align}
	where $\alpha=x, y, z$ is the direction of the spin polarization, and $\mathcal{T}$ expresses the non-conservation processes, i.e., the source and sink of the spin due to the spin relaxation.
	In order to overcome these problems, there has been a extensive debate.
	For example, the definition was adjusted by separating the spin current into conserved and non-conserved parts.\cite{Murakami2004a}
	With this definition, at least for the spin current for the conserved part can be uniquely defined, but still the non-conserved part remains.

	Another approach is to introduce the torque dipole density.
	Shi {\it et al.} consider the $\mathcal{T}$ term in eq. (\ref{nonconserve}) to be the torque density $\mathcal{T}_z (\br) = {\rm Re} \psi^\dagger (\br) \hat{\tau} \psi (\br)$, where $\hat{\tau}\equiv d s_z / dt = -\Im \hbar^{-1}[s_z, \scr{H}]$.\cite{Shi2006}
	For the systems that the average spin torque density vanishes in the bulk, since the torque density can be expressed in terms of a divergence of a torque dipole density as $\mathcal{T}_z (\br)=-\bm{\nabla}\cdot \bm{P}_\tau (\br)$, one can write the continuity equation for the spin density and spin current in the form
	\begin{align}
		\frac{\partial \rho_{\rm s}^z}{\partial t}+\bm{\nabla}\cdot \left\{ \bm{j}_{\rm s}^z + \bm{P}_\tau (\br) \right\} = 0.
\end{align}
	This means that the spin density is conserved on average, and the corresponding transport current is
	\begin{align}
		\mathcal{J}_{\rm s} = \bm{j}_{\rm s} + \bm{P}_\tau.
\end{align}
	Based on this definition of spin current, the spin Hall conductivities are dramatically different from the conventional spin Hall conductivites.\cite{Shi2006,Sugimoto2006}

	The key of this difficult problem is the spin-orbit interaction, which is a relativistic effect.
	However, most of the discussions on this issue is based on a non-relativistic theory.
	Vernes {\it et al.} gave a fully relativistic theory of spin current and spin-transfer torque.\cite{Vernes2007}
	They proposed that the problem of the non-conserved spin current can be resolved by the choice of a convenient and covariant description of the spin polarization instead of the usual spin operators in the non-relativistic theory.
	They used the four component polarization operator $T_\mu \equiv (\bm{T}, T_4)$ introduced by Bargmann and Wigner\cite{Bargmann1948,Fradkin1961} to describe the spin polarization of moving electrons.
	$T_\mu$ commutes with the field-free Dirac Hamiltonian, so that the corresponding vector density satisfies a continuity equation.\cite{Lowitzer2010}

	The relativistic definition with $T_\mu$ seems to be the satisfactory one at the present moment in the sense that the spin-orbit coupling is treated in a fully relativistic way and the conserved spin current is defined from the continuity equation.
	However, there is an alternative to the polarization operator in the relativistic form.\cite{Fradkin1961}
	The different polarization operators lead to different results of spin Hall conductivity.
	The spin current cannot be evaluated uniquely even if we introduce the relativistic four component polarization operator after all.

	To summarize the present understanding of the spin current, there is no general and appropriate definition, even though we have several options for the conserved spin current.
	And even if we have the one, there still remains a problem whether the spin current actually corresponds to the observable physical quantity or not.\cite{Nomura2005c,Shi2006,Galitski2006}
	The most concrete and reliable way is to capture the experimental situation appropriately, and then calculate the corresponding observable physical quantities, such as the magnetization or the electric voltage.
	Nevertheless, the natural and simple definition as $j_\nu^\mu \equiv \frac{1}{2}\left\{ v_\nu , s_\mu \right\}$ will give a first brief idea for the SHE.
	%


\bibliographystyle{jpsj}
\bibliography{Bismuth,Bi_footnote}

\begin{thebibliography}{100}
\bibitem{note5}
By using the relations $C \psi_{n, \bm{k}, \uparrow} = \psi_{n, \bm{k},
  \downarrow}$, $\langle \phi | \psi \rangle = \langle C\psi | C \phi \rangle$,
  and $C^2 =-1$ we have \begin{align} \bm{t}&=\langle \psi_{{\rm c} \uparrow} |
  \bm{v} \psi_{{\rm v} \uparrow} \rangle =\langle C \bm{v} \psi_{{\rm v}
  \uparrow} | C \psi_{{\rm c} \uparrow} \rangle =\langle \bm{v} C\psi_{{\rm v}
  \uparrow} | \psi_{{\rm c} \downarrow} \rangle =\langle \psi_{{\rm v}
  \downarrow} | \bm{v} \psi_{{\rm c} \downarrow} \rangle, \\ \bm{u}&=\langle
  \psi_{{\rm c} \uparrow}| \bm{v} \psi_{{\rm v} \downarrow} \rangle =\langle
  \psi_{{\rm c} \uparrow} | \bm{v} C \psi_{{\rm v} \uparrow} \rangle =\langle
  C\bm{v} C \psi_{{\rm v} \uparrow} | C \psi_{{\rm c} \uparrow} \rangle
  =-\langle \bm{v} \psi_{{\rm v} \uparrow} | C \psi_{{\rm c} \uparrow} \rangle
  \nonumber\\ &=-\langle \psi_{{\rm v} \uparrow} | \bm{v} \psi_{{\rm c}
  \downarrow} \rangle, \end{align} where ${\rm c}$ and ${\rm v}$ denote
  conduction and valence band, respectively.
\bibitem{note8}
For a perfect crystal of $\alpha$-ET$_2$I$_3$, $\mu$ should locate just at the
  point, where the conduction band touches the valence band. However, $\mu$ can
  shift from this point if there is the lack of I$_3^-$ ions. Moreover, $\mu$
  can be temperature dependent since the conduction and valence bands are not
  perfectly symmetric. Thus $\mu$ can change sign at finite temperature $T_{\rm
  r}$ as is shown in Ref. \citen{Kobayashi2008}.
\bibitem{note6}
In semiconductors with zinc blend structure, such as GaAs, the Luttinger
  Hamiltonian is valid for the temperature higher than the energy of the
  splitting between the light hole and heavy hole bands, which is due to the
  lack of the inversion symmetry.
\end{thebibliography}

\end{document}